\documentclass[useAMS,usenatbib]{mn2e}  
\usepackage{graphicx,amsmath,amssymb,afterpage}
\newcommand \radm{rad~m$^{-2}$}
\begin{document}
   \title[Finding a faint polarized radio signal]{Finding a faint polarized signal in wide-band radio data}

   \author[D.H.F.M. Schnitzeler and K.J. Lee]
          {D.H.F.M. Schnitzeler$^{1}$\thanks{dschnitzeler@gmail.com} 
           and K.J. Lee$^{2}$\\ 
          $^1$ Max Planck Institut f\"ur Radioastronomie, D-53121 Bonn, Germany\\
          $^2$ Kavli Institute for Astronomy and Astrophysics, Peking University, Beijing 100871, People's Republic of China\\
          }

   \date{Accepted 2016 November 28. Received 2016 November 28; in original form 2016 July 19}

   \pagerange{}\pubyear{}

   \maketitle

\begin{abstract}
We develop two algorithms, based on maximum likelihood (ML) inference, for estimating the parameters of polarized radio sources which emit at a single rotation measure (RM), e.g., pulsars. 
These algorithms incorporate the flux density spectrum of the source, either a power law or a scaled version of the Stokes $I$ spectrum, and a variation in sensitivity across the observing band.
We quantify the detection significance and measurement uncertainties in the fitted parameters, and we derive weighted versions of the RM synthesis algorithm which, under certain conditions, maximize the likelihood.
We use Monte Carlo simulations to compare injected and recovered source parameters for a range of signal-to-noise ratios, investigate the quality of standard methods for estimating measurement uncertainties, and search for statistical biases.
These simulations consider one frequency band each for the Australia Telescope Compact Array (ATCA), the Square Kilometre Array (SKA), and the Low Frequency Array (LOFAR).
We find that results obtained for one frequency band cannot be easily generalized, and that methods which were developed in the past for correcting bias in individual frequency channels do not apply to wide-band data sets. 
The standard method for estimating the measurement uncertainty in RM is not accurate for sources with non-zero spectral indices.
Furthermore, dividing Stokes $Q$ and $U$ by Stokes $I$ to correct for spectral index effects,  in combination with RM synthesis, does not maximize the likelihood.
\end{abstract}

%
%________________________________________________________________

\section{Introduction}\label{introduction.sec}
Faraday rotation of polarized radio waves provides us with an important tool for studying magnetic fields in ionized gas, both locally in the Milky Way (e.g., \citealt{gardner1963} and \citealt{seielstad1964}) and out to very high redshifts (e.g., \citealt{carilli1994}). 
When astrophysical Faraday rotation was first discovered by \cite{cooper1962}, and in the following decades, measuring Faraday rotation required re-observing a source at a number of different radio frequencies.
With the advent of broad-band radio receivers it became possible to measure Faraday rotation with a single observation.
Nowadays the two most popular methods for extracting information from such observations are RM synthesis \citep{brentjens2005} and fitting models to measurements of Stokes $Q$ and $U$ as a function of frequency (`QU fitting'; e.g., \citealt{farnsworth2011} and \citealt{osullivan2012}).
Because measuring Faraday rotation requires studying the same source over a wide range of frequencies, two effects can complicate the interpretation of the data: 
radio sources generally have a non-zero flux density spectral index, and the sensitivity of the observations can vary between frequencies. 
Such a variation in the sensitivity of the observations can arise, for example, because the system temperature varies across the observing band, or because strong radio frequency interference makes data points unusable.
Several authors have tried to mitigate the first effect by dividing the observed Stokes $Q$ and $U$ flux densities by the observed Stokes $I$ flux density, and the second effect by using one over the measured noise variance of each frequency channel as statistical weights. 
As far as we know, no mathematical proof for the latter method being correct has been published previously.
Fitting the Stokes $Q$ and $U$ frequency spectra by maximizing the likelihood can solve both problems: one can fit for the polarized flux density spectrum of the source and use measurements of the noise variances in Stokes $Q$ and $U$ to incorporate a variation in sensitivity at different frequencies.

In this paper we will use semi-analytical techniques to determine the polarization properties of simple sources that emit at only one rotation measure RM, 
\begin{eqnarray}
\mathrm{RM}\, \left(\mathrm{rad~m}^{-2}\right)\ \approx\ 0.81 \int_\mathrm{source}^\mathrm{observer} n_\mathrm{e} B_\| \mathrm{d}l\, .
\label{rm_definition.eqn}
\end{eqnarray}
Here $n_\mathrm{e}$ is the free electron density (cm$^{-3}$), $B_\|$ the length of the magnetic field vector projected along the line of sight ($\mu$G), and d$l$ an infinitesimal distance interval along the line of sight from the source to the observer (pc).
We will refer to the integral in equation~\ref{rm_definition.eqn} as the RM of the emission, instead of the Faraday depth of the emission. In Appendix~\ref{nomenclature.sec} we argue why we prefer this nomenclature. 
In our paper we will quantify the measurement uncertainties of the parameters that describe the radio signal and its detection significance.
In two cases we were able to derive analytical expressions for maximizing the likelihood, which greatly reduces the dimensionality of the search grid, thereby speeding up data processing.
These analytical expressions shed light on the connection between the techniques of 
QU fitting and RM synthesis, and can be used as simple and fast tests for determining whether a source emits at a single RM. If the source fails this test, a more complex model might be required to describe the source.
This is not the first time ML techniques have been used to determine the (polarization) properties of radio sources:
\cite{osullivan2012}, \cite{scaife2012}, and \cite{wehus2013} used brute force methods to search through parameter space, while \cite{bell2013} used ML inference to improve the cleaning of RM spectra (see also section~2.2 in \citealt{sun2015}).
\cite{montier2015b} discuss additional methods for estimating parameters of polarized radio sources whose emission is not affected by Faraday rotation.
Finally, this paper provides background information on the algorithm that we used to determine the polarization properties of radio pulsars  in the Galactic Centre (\citealt{schnitzeler2016}, `S16').

This paper is structured as follows. 
In Section~\ref{parameters.sec}, we derive ML estimators for a radio source whose polarized flux density spectrum is either a power law or a scaled version of the Stokes $I$ spectrum.
We also derive a weighted form of RM synthesis which maximizes the likelihood for weakly polarized sources, and we show that using ratios of measured flux densities $Q/I$ and $U/I$ in RM synthesis does not maximize the likelihood.
In Section~\ref{uncertainties.sec} we use Monte Carlo simulations to investigate parameter distributions in the presence of noise, we test the accuracy of standard methods for calculating measurement uncertainties, and we look for potential bias in the estimators we derived.

Throughout our paper we will use $\bmath{L}\, =\, Q+\mathrm{i}\,U$ to indicate the linear polarization vector, as this variable name is commonly used instead of $\bmath{P}$ in pulsar observations.
Furthermore, we will ignore the fact that frequency channels have a finite width, and that channel weighting functions should be included when considering sources with large $|\mathrm{RM}|$ (as we showed in \citealt{schnitzeler2015}) or with large variations in flux density across the frequency channels.

\section{Parameter estimation}\label{parameters.sec}
The problem that we aim to solve can be summarized as follows.
Given measurements of the Stokes parameters $Q$ and $U$ and their noise variances at various frequencies, find the parameters which describe the polarized radio signal: 
the intrinsic Stokes parameters $Q_\mathrm{ref}$ and $U_\mathrm{ref}$ at a specific reference frequency, the shape of the polarized flux density spectrum, and the RM of the signal.
We will indicate the measured Stokes $Q$ and $U$ flux densities with $Q_{\mathrm{obs}}$ and $U_{\mathrm{obs}}$, and we will assume that measurements are available for $N_\mathrm{ch}$ mutually independent frequency channels\footnote{\cite{montier2015a} include the full covariance matrix for the Stokes parameters $I, Q$, and $U$ in their analysis of the polarization fraction and polarization angle of radio sources. However, the sources they consider are not affected by Faraday rotation.}.
Furthermore, we assume that there are no offsets in $Q$ and $U$ across the frequency band, and that the noise in each of the channels follows a Gaussian distribution with variance $\sigma_{Q,i}^2$ and $\sigma_{U,i}^2$ for Stokes $Q$ and $U$, where $i$ is the channel index. Different frequency channels can have different noise variances, and the noise variances in Stokes $Q$ and $U$ do not have to be the same. 
We include a factor $\eta$ which corrects for a possible error in the scale of $\sigma_{Q,i}$ and $\sigma_{U,i}$. The log likelihood for this observing setup is given by
\begin{eqnarray}
\log{\Lambda} = 
-\frac{1}{2}\sum_{i=1}^{N_\mathrm{ch}}
          \left[\left(\frac{Q_{\mathrm{mod},i}\,c_i - U_{\mathrm{mod},i}\,s_i - Q_{\mathrm{obs},i}}{\eta\,\sigma_{Q,i}}\right)^2 \right. \nonumber \\
+ \left.\left(\frac{Q_{\mathrm{mod},i}\,s_i + U_{\mathrm{mod},i}\,c_i - U_{\mathrm{obs},i}}{\eta\,\sigma_{U,i}}\right)^2 \right] \nonumber \\
- \sum_{i=1}^{N_\mathrm{ch}} \left[\log\left(\sigma_{Q,i}\right) + \log\left(\sigma_{U,i}\right)\right] \nonumber \\
-N_\mathrm{ch}\left[\log\left(2\pi\right)+2\log\left(\eta\right)\right]\, .
\label{loglikelihood.eqn}
\end{eqnarray}
$Q_{\mathrm{mod},i}$ and $U_{\mathrm{mod},i}$ describe the emitted (intrinsic) polarized flux density spectrum of the source. Faraday rotation of this signal is described by $c_i$ and $s_i$,
which are shorthand for $\cos{\left(2\mathrm{RM}\lambda^2\right)}$ and $\sin{\left(2\mathrm{RM}\lambda^2\right)}$, respectively. $\lambda^2$ is the mean wavelength squared of a frequency channel.
When searching for sources with large positive or negative RMs, or with steep flux density spectra, the four possible combinations of $\left(Q_{\mathrm{mod},i}, U_{\mathrm{mod},i}\right)\times\left(c_i, s_i\right)$ in equation~\ref{loglikelihood.eqn} should be replaced with their channel-averaged values (see also \citealt{schnitzeler2015}).

Equation~\ref{loglikelihood.eqn} requires solving for the $Q$ and $U$ flux density of each frequency channel, and for RM and $\eta$, which means that this system of equations is underdetermined. 
To reduce the dimensionality of the parameter space we consider two special cases: sources with a power-law polarized flux density spectrum (Section~\ref{powerlaw_spectrum.sec}), and sources for which the polarized flux density spectrum is a scaled version of the Stokes $I$ spectrum (Section~\ref{scaled_spectrum.sec}). 
In Section~\ref{ML_RMsynthesis.sec} we derive expressions for RM synthesis that allow for a variation in the polarized flux density spectrum of the source and for a variation in sensitivity across the observing band, and we show under which conditions RM synthesis maximizes the likelihood.
Finally, in Section~\ref{Detection_probability.sec} we discuss how the likelihood can be used to quantify the significance of the detection, i.e., whether the signal can be explained as purely due to noise.

Correcting for spectral index effects by dividing Stokes $Q$ and $U$ by Stokes $I$ requires (amongst others) that linear polarization and Stokes $I$ are produced in the same part of the source, which is not guaranteed to be the case. For example, when studying active galactic nuclei the core can dominate the emission in Stokes $I$ but show very little linear polarization (e.g. in gigahertz-peaked spectrum sources, \citealt{odea1998}). In such cases it makes little sense to divide Stokes $Q$ and $U$ by Stokes $I$.

\subsection{Power-law polarized flux density spectrum}\label{powerlaw_spectrum.sec}
In this case the polarized flux density spectrum is a simple power law, and the linear Stokes parameters can be written as
$Q_{\mathrm{mod},i} = Q_\mathrm{ref}\left(\nu_i/\nu_\mathrm{ref}\right)^\alpha$ and $U_{\mathrm{mod},i} = U_\mathrm{ref}\left(\nu_i/\nu_\mathrm{ref}\right)^\alpha$.
The ML estimators for $Q_\mathrm{ref}, U_\mathrm{ref}$, and $\eta$, are found by taking the derivative of the log likelihood in equation~\ref{loglikelihood.eqn} with respect to these parameters and setting the result equal to zero. 
We will indicate ML estimators by using hats, for example in $\hat{Q}_\mathrm{ref}$ and $\hat{U}_\mathrm{ref}$.
The equations that maximize the likelihood for Stokes $Q_\mathrm{ref}$ and $U_\mathrm{ref}$ can be written in matrix form as
\begin{eqnarray}
\lefteqn{
\begin{pmatrix} a_{1,1} & a_{1,2} \\ a_{1,2} & a_{2,2}\\
\end{pmatrix}
\begin{pmatrix} \hat{Q}_\mathrm{ref} \\ \hat{U}_\mathrm{ref}\\
\end{pmatrix}
= } \nonumber\\
& & 
\sum_{i=1}^{N_\mathrm{ch}} \left(\frac{\nu_i}{\nu_\mathrm{ref}}\right)^{\alpha}
\begin{pmatrix}
c_i & s_i \\
-s_i & c_i\\
\end{pmatrix}
\begin{pmatrix}
Q_{\mathrm{obs},i}/\sigma_{Q,i}^2 \\ U_{\mathrm{obs},i}/\sigma_{U,i}^2 \\
\end{pmatrix}
\end{eqnarray}
where
\begin{eqnarray}
a_{1,1} & = & \sum_{i=1}^{N_\mathrm{ch}} \left(\frac{\nu_i}{\nu_\mathrm{ref}}\right)^{2\alpha}\left(\frac{c_i^2}{\sigma_{Q,i}^2} + \frac{s_i^2}{\sigma_{U,i}^2}\right) \nonumber\\
a_{1,2} & = & \sum_{i=1}^{N_\mathrm{ch}} \left(\frac{\nu_i}{\nu_\mathrm{ref}}\right)^{2\alpha}\left(-\frac{c_i\,s_i}{\sigma_{Q,i}^2} + \frac{s_i\,c_i}{\sigma_{U,i}^2}\right) \nonumber\\
a_{2,2} & = & \sum_{i=1}^{N_\mathrm{ch}} \left(\frac{\nu_i}{\nu_\mathrm{ref}}\right)^{2\alpha}\left(\frac{s_i^2}{\sigma_{Q,i}^2} + \frac{c_i^2}{\sigma_{U,i}^2}\right) \nonumber
\nonumber
\end{eqnarray}
The ML estimators for $Q_\mathrm{ref}$ and $U_\mathrm{ref}$ can then be found by matrix inversion, and the square of the ML estimator for $\eta$ is 
\begin{eqnarray}
\lefteqn{
\left(\hat{\eta}\right)^2 = 
\frac{1}{2N_\mathrm{ch}}\sum_{i=1}^{N_\mathrm{ch}} 
\left[
\left(\frac{\nu_i}{\nu_\mathrm{ref}}\right)^{2\alpha}
  \left(\frac{\hat{Q}_\mathrm{ref}\,c_i - \hat{U}_\mathrm{ref}\,s_i - Q_{\mathrm{obs},i}}{\sigma_{Q,i}}\right)^2 \right.
} \nonumber \\
 & & + \left. \left(\frac{\hat{Q}_\mathrm{ref}\,s_i + \hat{U}_\mathrm{ref}\,c_i - U_{\mathrm{obs},i}}{\sigma_{U,i}}\right)^2 
 \right]\, .
\end{eqnarray}
Because the equations for $\hat{Q}_\mathrm{ref}, \hat{U}_\mathrm{ref}$ and $\hat{\eta}$ can be written in terms of $\alpha$ and RM, maximising the likelihood requires searching through only a two-dimensional grid (RM, $\alpha$) instead of the five-dimensional grid (RM, $\alpha$, $Q_\mathrm{ref}, U_\mathrm{ref}$, and $\eta$).

\subsection{Stokes $L$ is a scaled version of Stokes $I$}\label{scaled_spectrum.sec}
If the polarized flux density spectrum is a scaled version of the Stokes $I$ spectrum (which is usually the case in simple polarized sources) then $Q_{\mathrm{mod},i}\,= q\,I_{\mathrm{mod},i}$ and $U_{\mathrm{mod},i}\,= u\,I_{\mathrm{mod},i}$; $q$ and $u$ describe the intrinsic linear polarization properties of the source, and do not depend on frequency. 
If instead we make the assumption that the \emph{measured} flux densities in Stokes $Q$ and $U$ are proportional to Stokes $I$ then this implies also a correlation between the noise contributions to these Stokes parameters, which is contrary to our assumptions.
Assuming Gaussian noise in Stokes $I$ with a variance $\sigma_{I,i}^2$, independent frequency channels, and no correlation between the Stokes parameters, including the observed Stokes $I$ spectrum in the log likelihood adds an extra term 
\begin{eqnarray}
-\frac{1}{2}\sum_{i=1}^{N_\mathrm{ch}}\left(\frac{I_{\mathrm{mod},i}-I_{\mathrm{obs},i}}{\sigma_{I,i}}\right)^2 - \sum_{i=1}^{N_\mathrm{ch}}\log{\sigma_{I,i}} 
- \frac{N_\mathrm{ch}}{2}\log{\left(2\pi\right)} 
\end{eqnarray}
to equation~\ref{loglikelihood.eqn}. 
To simplify our analysis we will assume that the measured noise variances in Stokes $I$ are exact, and do not require a scale factor $\eta_I$.
Since we are searching for ways to assign weights to individual frequency channels in RM synthesis based on the noise variance of each channel and the source spectral index, we will only consider algorithms that search through a 1D grid of trial RM values, similar to RM synthesis.

In appendix~\ref{Appendix_A} we derive the equations for the ML estimators of $q$ and $u$. 
These equations involve polynomials of degree six or seven, depending on how the `nuisance parameters' $I_{\mathrm{mod,i}}$ are removed. 
There are no general solutions for such polynomials with arbitrary coefficients (based on the Abel-Ruffini theorem, \citealt{jacobson2009}), and numerical methods have to be used to maximize the likelihood.
Furthermore, sums over all frequency channels of the ratios $Q_{\mathrm{obs},i}/I_{\mathrm{obs},i}$ and $U_{\mathrm{obs},i}/I_{\mathrm{obs},i}$ do not occur in the equations for finding the ML estimators of $q$ and $u$. 
Therefore, using these flux density ratios in RM synthesis to correct for spectral index effects does not lead to the ML estimators for $q$ and $u$, which is undesirable.
This result can be understood intuitively. 
Plotting $Q_{\mathrm{obs},i}$ and $U_{\mathrm{obs},i}$ in the complex plane will show a scatter plot arranged along a segment of a spiral which can be extrapolated back to the origin (if $\alpha=0$ all data points scatter along a circle segment, and if $\mathrm{RM}=0$ the spiral becomes a radial line).
Now consider plotting $Q_{\mathrm{obs},i}/I_{\mathrm{obs},i}$ and $U_{\mathrm{obs},i}/I_{\mathrm{obs},i}$ in the complex plane, and compare the two figures.
If the signal-to-noise level of the data is poor, dividing $Q_{\mathrm{obs},i}$ and $U_{\mathrm{obs},i}$ by $I_{\mathrm{obs},i}$ means dividing two noisy quantities by another noisy quantity. 
The noise in $I_{\mathrm{obs},i}$ reshuffles many data points over the four quadrants, and one cannot hope to recover the correct (ML) estimators for $q$ and $u$ in this case.
Perhaps at a sufficiently high signal-to-noise level (small scatter in the complex plane) applying RM synthesis to $Q_{\mathrm{obs},i}/I_{\mathrm{obs},i}$ and $U_{\mathrm{obs},i}/I_{\mathrm{obs},i}$ can approximate maximizing the likelihood, but we did not explore this possibility further.

Fortunately, under certain conditions the log likelihood we derived after marginalizing over $I_{\mathrm{mod,i}}$ (equation~\ref{loglikelihood_prime.eqn}) can be simplified to
\begin{eqnarray}
\log{\Lambda'} & = & 
-\frac{1}{2}\sum_{i=1}^{N_\mathrm{ch}}
          \left[\left(\frac{Q_{\mathrm{obs},i} - \alpha_i\,I_{\mathrm{obs},i}}{\eta\,\sigma_{L,i}}\right)^2 \right. \nonumber \\
 & & \left. + \left(\frac{U_{\mathrm{obs},i} - \beta_i\,I_{\mathrm{obs},i}}{\eta\,\sigma_{L,i}}\right)^2\right. \nonumber \\
 & & \left. + \left(\frac{\beta_i\,Q_{\mathrm{obs},i} - \alpha_i\,U_{\mathrm{obs},i}}{\eta\,\sigma_{L,i}}\right)^2
 \right] \nonumber \\
 & & - 2\sum_{i=1}^{N_\mathrm{ch}} \log\left(\sigma_{L,i}\right) 
       -N_\mathrm{ch}\left[\log\left(2\pi\right)+2\log\left(\eta\right)\right]
\label{loglikelihood_prime_simplified.eqn}
\end{eqnarray}
(see appendix~\ref{Appendix_B} for the derivation; $\alpha_i$ and $\beta_i$ are defined in equation~\ref{loglikelihood_prime.eqn}).
This equation is valid if in addition to $L \propto I$ also the noise variances in Stokes $Q$ and $U$ are equal ($\sigma_{L,i}^2 \equiv\sigma_{Q,i}^2 = \sigma_{U,i}^2$), the source is weakly polarized ($L_{\mathrm{mod}} \ll I_{\mathrm{mod}}$), and $\sigma_{I,i}~\approx~\eta\,\sigma_{L,i}$.

The only free parameters in equation~\ref{loglikelihood_prime_simplified.eqn} are $q, u$, RM, and $\eta$, RM being the only non-linear parameter.
The square of the ML estimator for $\eta$ is 
\begin{eqnarray}
\lefteqn{
\left(\hat{\eta}\right)^2 = 
\frac{1}{2N_\mathrm{ch}}\sum_{i=1}^{N_\mathrm{ch}} 
 \frac{1}{\sigma_{L,i}^2} \times
} \nonumber\\
 & & 
 \left[ \left( Q_{\mathrm{obs},i} - \left(c_i \hat{q} - s_i \hat{u}\right)\,I_{\mathrm{obs},i} \right)^2  \right. \nonumber \\
 & & \left. 
 + \left(U_{\mathrm{obs},i} - \left(s_i \hat{q} + c_i \hat{u}\right)\,I_{\mathrm{obs},i} \right)^2 \right. \nonumber \\
 & & \left. 
 + \left(\hat{q}\left(s_i\,Q_{\mathrm{obs},i} - c_i\,U_{\mathrm{obs},i}\right) + 
 \hat{u}\left(c_i\,Q_{\mathrm{obs},i} + s_i\,U_{\mathrm{obs},i}\right)\right)^2 
\right]\, . \nonumber\\
\end{eqnarray}
In this case the expressions for the ML estimators for $q, u$, and $\eta$ can be written as a function of RM, which means that the (slow) 2D grid search from Section~\ref{powerlaw_spectrum.sec} is reduced to a much faster 1D grid search over RM.
This is where our ML-based method starts to show similarities with RM synthesis, and we will discuss this in more detail in Section~\ref{ML_RMsynthesis.sec}.

\subsection{ML optimization and RM synthesis}\label{ML_RMsynthesis.sec}
\subsubsection{Power-law polarized flux density spectrum}
If the source has a power-law flux density spectrum, all noise variances in $Q$ and $U$ are equal ($\sigma_{Q,i}^2 = \sigma_{U,i}^2\equiv\sigma_{L,i}^2$), and one searches the (RM,$\alpha$) grid only along the RM axis at $\alpha =0$ then
\begin{eqnarray}
\begin{pmatrix} \hat{Q}_\mathrm{ref} \\ \hat{U}_\mathrm{ref} \\
\end{pmatrix} = 
\sum_{i=1}^{N_\mathrm{ch}}\frac{1}{\sigma_{L,i}^2}
\begin{pmatrix}
c_i & s_i \\
-s_i & c_i\\
\end{pmatrix}
\begin{pmatrix} Q_{\mathrm{obs},i} \\ U_{\mathrm{obs},i} \\
\end{pmatrix}
/\sum_{i=1}^{N_\mathrm{ch}}\frac{1}{\sigma_{L,i}^2}\, .
\label{ml_rmsynthesis_powerlaw.eqn}
\end{eqnarray}
In particular, if the noise variances do not depend on frequency 
\begin{eqnarray}
\begin{pmatrix} \hat{Q}_\mathrm{ref} \\ \hat{U}_\mathrm{ref} \\
\end{pmatrix} = 
\frac{1}{N_\mathrm{ch}}\sum_{i=1}^{N_\mathrm{ch}}
\begin{pmatrix}
c_i & s_i \\
-s_i & c_i\\
\end{pmatrix}
\begin{pmatrix} Q_{\mathrm{obs},i} \\ U_{\mathrm{obs},i} \\
\end{pmatrix}\, .
\label{ml_rmsynthesis.eqn}
\end{eqnarray}
The right-hand side of equation~\ref{ml_rmsynthesis_powerlaw.eqn} expresses how the observed polarization vectors $\left(Q_{\mathrm{obs},i}, U_{\mathrm{obs},i}\right)$ are derotated and summed, using the noise variances as weights. 
If all noise variances are equal one recognizes the right-hand side of equation~\ref{ml_rmsynthesis.eqn} as the net polarization vector which is calculated using RM synthesis.
Summarizing, if the source has a power-law spectrum with a spectral index of zero, RM synthesis produces the ML estimators of $Q_\mathrm{ref}$ and $U_\mathrm{ref}$; if the noise variances vary across the frequency band then equation~\ref{ml_rmsynthesis_powerlaw.eqn} can be used to include this variation to calculate the ML estimators for $Q_\mathrm{ref}$ and $U_\mathrm{ref}$.

\begin{figure}
\resizebox{\hsize}{!}{\includegraphics{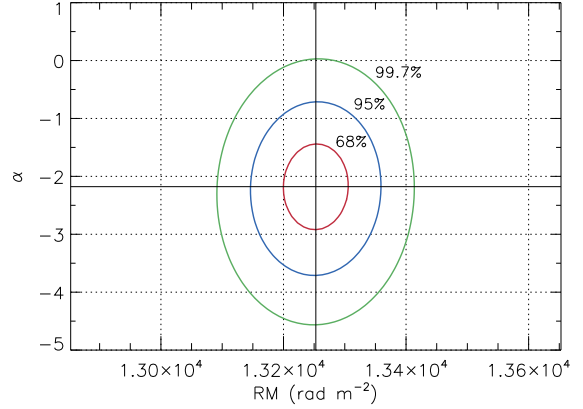}}
\caption{Log likelihood contours for the observations of PSR J1746-2856 reported in S16. The red, blue, and green contours indicate the 68\%, 95\%, and 99.7\% confidence intervals for one source parameter, respectively. 
The grid which we show in this figure is intended to help guide the eye; the grid we actually used for calculating the contour levels is much finer. The axis ratio of the contour ellipses depends on the frequency coverage of the observations.
}
\label{rm_alpha_contours.fig}
\end{figure}

If RM synthesis (or its weighted form: equation~\ref{ml_rmsynthesis_powerlaw.eqn}) is applied to a source with a non-zero spectral index then this can affect the derived $Q_\mathrm{ref}$ and $U_\mathrm{ref}$, the significance of the detection, and the confidence intervals in the source parameters. 
For example, based on our polarization observations of PSR J1746-2856 with the ATCA between 4.5-6.5 GHz (shown in fig.~2 from S16) we concluded that this pulsar has a polarization spectral index which is significantly different from zero (table~1 from S16).
These observations are sensitive enough that the contour levels in the (RM,$\alpha$) grid are compact, and the log likelihood reaches its maximum far from the line $\alpha=0$ in the (RM,$\alpha$) grid (Fig.~\ref{rm_alpha_contours.fig}).
If the source is detected at a sufficiently high signal-to-noise level then the axes of the contour ellipses in the log likelihood landscape tend to line up with the (RM,$\alpha$) axes of the coordinate grid. 
Under these conditions it is possible that the correct RM and its associated measurement uncertainty can be found by applying RM synthesis even if the source has a non-zero spectral index.

\subsubsection{Stokes $L$ is a scaled version of Stokes $I$}
We find the ML estimators for $q$ and $u$ by taking the derivative of equation~\ref{loglikelihood_prime_simplified.eqn} with respect to these parameters and setting the resulting expressions equal to zero.
Re-arranging the expressions for $\partial \log\Lambda'/\partial{\{q,u\}}$ gives the following equation:
\begin{eqnarray}
\lefteqn{
\begin{pmatrix} 
k_{1,1} & k_{1,2} \\
k_{1,2} & k_{2,2} \\
\end{pmatrix}
\begin{pmatrix} \hat{q} \\ \hat{u} \\
\end{pmatrix} = } \nonumber\\
& & \sum_{i=1}^{N_\mathrm{ch}}\frac{I^2_{\mathrm{obs},i}}{\sigma_{L,i}^2}
\begin{pmatrix}
c_i & s_i \\
-s_i & c_i\\
\end{pmatrix}
\begin{pmatrix} Q_{\mathrm{obs},i}/I_{\mathrm{obs},i} \\ U_{\mathrm{obs},i}/I_{\mathrm{obs},i} \\
\end{pmatrix}
/\sum_{i=1}^{N_\mathrm{ch}}\frac{I^2_{\mathrm{obs},i}}{\sigma_{L,i}^2}
\label{ml_rmsynthesis_scaled.eqn}
\end{eqnarray}
The matrix on the left-hand side of equation~\ref{ml_rmsynthesis_scaled.eqn} has the following elements:
\begin{eqnarray}
k_{1,1} & = & 1+ \sum_{i=1}^{N_\mathrm{ch}} \frac{\left(Q_{\mathrm{obs},i}\, s_i - U_{\mathrm{obs},i}\, c_i\right)^2}{\sigma_{L,i}^2}
/\sum_{i=1}^{N_\mathrm{ch}}\frac{I^2_{\mathrm{obs},i}}{\sigma_{L,i}^2}  
\\
k_{1,2} & = & \nonumber \\
\lefteqn{\sum_{i=1}^{N_\mathrm{ch}} \frac{\left(Q_{\mathrm{obs},i}c_i + U_{\mathrm{obs},i}s_i\right)\left(Q_{\mathrm{obs},i}s_i - U_{\mathrm{obs},i}c_i\right)}{\sigma_{L,i}^2}
/\sum_{i=1}^{N_\mathrm{ch}}\frac{I^2_{\mathrm{obs},i}}{\sigma_{L,i}^2}}
\nonumber \\
& & \\
k_{2,2} & = & 1+ \sum_{i=1}^{N_\mathrm{ch}} \frac{\left(Q_{\mathrm{obs},i}\, c_i + U_{\mathrm{obs},i}\, s_i\right)^2}{\sigma_{L,i}^2}
/\sum_{i=1}^{N_\mathrm{ch}}\frac{I^2_{\mathrm{obs},i}}{\sigma_{L,i}^2}\, .
\end{eqnarray}
The right-hand side of equation~\ref{ml_rmsynthesis_scaled.eqn} expresses how the measured polarization vectors $\left(Q_{\mathrm{obs},i}, U_{\mathrm{obs},i}\right)$ are derotated and summed, similar to equation~\ref{ml_rmsynthesis_powerlaw.eqn} but with the weights 1/$\sigma_{L,i}^2$ replaced with $I^2_{\mathrm{obs},i}/\sigma_{L,i}^2$.
If all noise variances and all $I_{\mathrm{obs},i}$ are equal one recognizes the right-hand side of equation~\ref{ml_rmsynthesis_scaled.eqn} as the net polarization vector which is calculated with RM synthesis. 
As far as we know, this is the first time that an expression has been derived algebraically for RM synthesis that simultaneously allows for a variation in the polarized flux density spectrum of the source and for a variation in the sensitivity across the frequency band.

Inserting the solutions for $\hat{q}$ and $\hat{u}$ from equation~\ref{ml_rmsynthesis_scaled.eqn} back into equations~\ref{dlogLdq_margin.eqn} and \ref{dlogLdu_margin.eqn} shows that these $\hat{q}$ and $\hat{u}$ are not roots of the equations, and therefore do not maximize the likelihood.
We did not investigate further how large the polarization fraction can be for equation~\ref{ml_rmsynthesis_scaled.eqn} to be a useful approximation.
One can show that if the signal-to-noise ratio in all Stokes parameters tends to infinity then the $\hat{q}$ and $\hat{u}$ from equation~\ref{ml_rmsynthesis_scaled.eqn} are equal to the $q$ and $u$ of the injected signal, independent of the RM and polarization fraction of the source (polarization fractions of up to one are allowed).
Numerical methods can be used to find the maximum in the log likelihood landscape, instead of relying on the approximations which we used to derive equation~\ref{ml_rmsynthesis_scaled.eqn}. This approach has the additional advantage of being able to handle also more complex situations, e.g., variations of the noise variances across the observing band.

\subsection{Significance of the detection}\label{Detection_probability.sec}
Detection statistics in the RM domain have been investigated previously by \cite{george2012}, \cite{hales2012}, and \cite{macquart2012}.
Contrary to previous studies, we use the log likelihood also to calculate the detection significance. 
If $\log\Lambda_0$ is the log likelihood equivalent of equation~\ref{loglikelihood.eqn}, \ref{loglikelihood_prime.eqn}, or \ref{loglikelihood_prime_simplified.eqn}, if there is no signal but only noise, then $2\log{\left(\Lambda/\Lambda_0\right)}$ follows a $\chi^2$ distribution with a number of degrees of freedom equal to the number of parameters in $\Lambda$ minus the number of parameters in $\Lambda_0$ (\citealt{wilks1938}). For this theorem to be true $N_\mathrm{ch} \gg 1$, which is satisfied by wide-band receivers. The connection between the log likelihood ratio and the $\chi^2$ distribution makes it possible to assign to any signal a probability that a measurement is produced purely by noise. 
This method for calculating the detection significance from data in the frequency domain instead of the RM domain is both easier and more reliable, because data in different frequency channels are independent and because the noise properties of the individual channels are understood well.

As a caveat, the log likelihood ratio can deviate from a $\chi^2$ distribution, as pointed out by, e.g., \cite{chernoff1954} and \cite{protassov2002}. We did not investigate this further in our paper.

\section{Power-law polarized flux density spectra: measurement uncertainty and bias}\label{uncertainties.sec}
In this Section we will investigate the distribution of the maximum likelihood estimators for Stokes $Q_\mathrm{ref}$ and $U_\mathrm{ref}$, RM, and $\alpha$ using Monte Carlo simulations of the noise. 
We use standard expressions to calculate the polarized flux density and intrinsic polarization angle of the emission from the ML estimators for $Q_\mathrm{ref}$ and $U_\mathrm{ref}$:
\begin{eqnarray}
L_\mathrm{ref} & = & \sqrt{\hat{Q}_\mathrm{ref}^2+\hat{U}_\mathrm{ref}^2} \\
\chi_\mathrm{ref} & = & \frac{1}{2}\mathrm{atan}\left(\frac{\hat{U}_\mathrm{ref}}{\hat{Q}_\mathrm{ref}}\right)\, , 
\end{eqnarray}
where $\chi_\mathrm{ref}$ is calculated taking into account which quadrant in the $Q,U$ plane the polarization vector $Q_\mathrm{ref} + \mathrm{i}\,U_\mathrm{ref}$ is in. 
Note that $L_\mathrm{ref}$ and $\chi_\mathrm{ref}$ themselves are not ML estimators.
The subscript `ref' in `$\chi_\mathrm{ref}$' might be somewhat confusing. To clarify, $\chi_\mathrm{ref}$ is the intrinsic polarization angle, i.e., after correcting for Faraday rotation; `ref' refers to the reference frequency for the power law flux density spectrum (Section~\ref{powerlaw_spectrum.sec}).

In this Section we will test how accurate standard methods for estimating the measurement uncertainties are, and we will look for statistical bias\footnote{By measuring a physical quantity and the associated noise variance one can reconstruct a distribution of possible values for the parameter of interest.  
The aim of debiasing is to centre this distribution on the value of the injected signal; debiasing does not replace the distribution of possible values of the quantity of interest with a single, noise-free value.}. 
Although our focus will be on observations between 4.5-6.5 GHz, which we used in S16, we will also investigate two additional frequency bands using Monte Carlo simulations (Table~\ref{mc_runs.tab}).
First we explain how we estimate the measurement uncertainties (Section~\ref{errors.sec}), then we explain how we set up the Monte Carlo simulations (Section~\ref{MCsetup.sec}).
In subsequent sections we analyse the influence of noise on the parameters we are interested in. 
We summarize our main results in Section~\ref{summary.sec}.

\subsection{Calculating measurement errors for individual observations}\label{errors.sec}
If the signal-to-noise level is sufficiently high that confidence regions are simply connected then the 68\% confidence interval of each parameter can be found by determining where the log likelihood has decreased by 0.5 (\citealt{avni1976}, \citealt{cash1976}, \citealt{lampton1976}, and \citealt{avni1978}). 
At low signal-to-noise levels the log likelihood landscape can show multiple peaks above the contour at max($\log{\Lambda}$)-0.5, in which case this condition is violated; this makes the method we describe for deriving measurement uncertainties based on the shape of the log likelihood landscape unreliable at low signal-to-noise levels.
Close to the grid point with the global maximum in $\log{\Lambda}$ we fitted a 2D parabola to accurately determine the ML estimators for RM and $\alpha$, and with these, $\hat{Q}_\mathrm{ref}$, $\hat{U}_\mathrm{ref}$, and $\hat{\eta}$.
To determine the uncertainties in RM and $\alpha$ we used the projection of the 2D ellipse where the likelihood has decreased by 0.5 onto the coordinate axes (e.g., figure 15.6.4 in \citealt{press1992}). 
Fig.~\ref{rm_alpha_contours.fig} illustrates this for PSR J1746-2856; also other significance levels are shown.
The ML estimators for $Q_\mathrm{ref}$ and $U_\mathrm{ref}$ depend on RM and $\alpha$, and we derived their measurement uncertainties by fitting a parabola to the log likelihood, varying only the parameter of interest and keeping the other parameters fixed at the values that maximize the likelihood.
This is the method we used in S16, and will be using in the current paper.

Alternatively, for a power-law polarized flux density spectrum we can calculate confidence intervals for $\hat{Q}_\mathrm{ref}, \hat{U}_\mathrm{ref}$, and $\hat{\eta}$ more quickly from the covariance matrix $\mathbfss{C}$ that is the inverse of the Fisher information matrix, $\mathbfss{I}$. 
$\mathbfss{I}$ has elements
\begin{eqnarray}
\mathbfss{I}_{i,j} = \Big<\frac{\upartial\log{\Lambda}}{\upartial\gamma_i}\,\frac{\upartial\log{\Lambda}}{\upartial\gamma_j}\Big>\, ,
\end{eqnarray}
where brackets indicate calculating the expectation value over the parameter space, using the likelihood as probability density function, and $\gamma_{i}$ indicates either $Q_\mathrm{ref}$, $U_\mathrm{ref}$, or $\eta$. 
Since the deviations between the model and the observations in equation~(2) 
are Gaussian, the matrix elements $\mathbfss{I}_{i,j}$ can be calculated using the Slepian-Bangs formula (\citealt{slepian1954}, \citealt{bangs1971}), 
which leads to
\begin{equation}
\mathbfss{C} = 
\begin{pmatrix}
\kappa \bmath{\delta}_2^\mathrm{T}\mathbfss{D}^{-1} \bmath{\delta}_2 & -\kappa \bmath{\delta}_1^\mathrm{T}\mathbfss{D}^{-1} \bmath{\delta}_2 & 0 \\
-\kappa  \bmath{\delta}_2^\mathrm{T}\mathbfss{D}^{-1} \bmath{\delta}_1 & \kappa \bmath{\delta}_1^\mathrm{T}\mathbfss{D}^{-1} \bmath{\delta}_1 & 0 \\
 0 & 0 & \hat{\eta}^2/4N_\mathrm{ch} \\
\end{pmatrix}
\end{equation}
Here,
\begin{eqnarray}
\bmath{\delta}_1^\mathrm{T} & = & 
  \left[
   \left(\frac{\nu_1}{\nu_\mathrm{ref}}\right)^{\hat{\alpha}}\, c_1, 
   \ldots ,
   \left(\frac{\nu_{N_\mathrm{ch}}}{\nu_\mathrm{ref}}\right)^{\hat{\alpha}}\, c_{N_\mathrm{ch}}, \right. \nonumber \\ 
   & & \left. \left(\frac{\nu_1}{\nu_\mathrm{ref}}\right)^{\hat{\alpha}}\, s_1,
   \ldots ,
   \left(\frac{\nu_{N_\mathrm{ch}}}{\nu_\mathrm{ref}}\right)^{\hat{\alpha}}\, s_{N_\mathrm{ch}}
  \right]\, , \nonumber
  \\
\bmath{\delta}_2^\mathrm{T} & = & 
  \left[
   - \left(\frac{\nu_1}{\nu_\mathrm{ref}}\right)^{\hat{\alpha}}\, s_1, 
   \ldots ,
   - \left(\frac{\nu_{N_\mathrm{ch}}}{\nu_\mathrm{ref}}\right)^{\hat{\alpha}}\, s_{N_\mathrm{ch}}, \right. \nonumber \\ 
   & & \left. \left(\frac{\nu_1}{\nu_\mathrm{ref}}\right)^{\hat{\alpha}}\, c_1,
   \ldots ,
   \left(\frac{\nu_{N_\mathrm{ch}}}{\nu_\mathrm{ref}}\right)^{\hat{\alpha}}\, c_{N_\mathrm{ch}}
  \right]\, , \nonumber
\end{eqnarray}
the matrix 
\begin{equation}
\mathbfss{D} = \hat{\eta}^2
\begin{pmatrix} \sigma_{Q,1}^2 \\
 & \ddots \\
 & 	& \sigma_{Q,N_\mathrm{ch}}^2 \\
 &	&	& \sigma_{U,1}^2 \\
 &	&	&	& \ddots \\
 &	&	&	&	& \sigma_{U,N_\mathrm{ch}}^2 \\
\end{pmatrix}
\end{equation}
(all off-diagonal elements being 0), and the scalar $\kappa = 1/\left(\bmath{\delta}_1^\mathrm{T}\mathbfss{D}^{-1} \bmath{\delta}_1\,\times\,\bmath{\delta}_2^\mathrm{T}\mathbfss{D}^{-1} \bmath{\delta}_2 - \bmath{\delta}_1^\mathrm{T}\mathbfss{D}^{-1} \bmath{\delta}_2\,\times\,\bmath{\delta}_2^\mathrm{T}\mathbfss{D}^{-1} \bmath{\delta}_1\right)$.
If the noise variances in Stokes $Q$ and $U$ are equal in all channels, but are allowed to vary between channels, i.e., $\sigma_{Q,i}^2=\sigma_{U,i}^2 \equiv \sigma_{L,i}^2$, then 
\begin{eqnarray}
\mathrm{var}\left(\hat{Q}_\mathrm{ref}\right) = \mathrm{var}\left(\hat{U}_\mathrm{ref}\right) = \hat{\eta}^2 \bigg/\sum_{i=1}^{N_\mathrm{ch}} \frac{1}{\sigma_{L,i}^2}\left(\frac{\nu_i}{\nu_\mathrm{ref}}\right)^{2\hat{\alpha}}\, . \nonumber
\end{eqnarray}
and cov($\hat{Q}_\mathrm{ref},\hat{U}_\mathrm{ref}$) = cov($\hat{U}_\mathrm{ref},\hat{Q}_\mathrm{ref}$) = 0. Furthermore, if $\hat{\eta} = 1$, $\hat{\alpha} = 0$, and all channels have the same noise variance then var$\left(\hat{Q}_\mathrm{ref}\right)$ = var$\left(\hat{U}_\mathrm{ref}\right) = \sigma^2/N_\mathrm{ch}$. 
Starting by assuming a constant noise variance for $N_\mathrm{ch}$ independent channels, \citet{macquart2012} derived the same expression for the noise level in the amplitude of the RM spectrum.

One can calculate the Fisher information also for equation~\ref{loglikelihood_prime_simplified.eqn}, but this is very time consuming. In this case fitting parabolas to the log likelihood is an alternative, practical solution for estimating the measurement uncertainties in $q, u$, and $\eta$.

\begin{table*}
\centering
\caption{Overview of our Monte Carlo simulations. The injected signal has an RM of 0~\radm, an intrinsic polarization angle $\chi_0 = 0$\degr, and a spectral index of 0, -2, or 2 (`a'/`d', `b', and `c' runs, respectively). 
`Telescope' lists an existing or planned telescope where this frequency coverage is available.
`FWHM RMSF' indicates the full width at half-maximum of the RM spread function.
RM$_\mathrm{max}$ is defined as the $|$RM$|$ where the recovered polarized flux density of a source has dropped to half the emitted flux density; see also footnote~1 in S16. To calculate this quantity we used \protect\cite{schnitzeler2015} to simulate frequency channels with a finite width and uniform channel response function. 
}
\label{mc_runs.tab}
\begin{tabular}{lcr@{-}lccccc}
\hline
	& 	Telescope	&	\multicolumn{2}{c}{Frequency} 	&	Channel	&	Channel	&	FWHM		& RM$_\mathrm{max}$			& RM search\\
	& 			&	\multicolumn{2}{c}{range}	&	number	&	width	&	RMSF		& 				& range\\
	&			&	\multicolumn{2}{c}{(MHz)}	&		&	(MHz)	&	(rad~m$^{-2}$)	& (rad~m$^{-2}$)		& (rad~m$^{-2}$)\\
\hline	
Run 1{a,b,c}	&	ATCA$^a$	&	4473.5 & 6525.5			&	513	&	4	&	$\approx$1600	& $\approx4.0\times10^5$	& $\pm8.0\times10^4$	\\
Run 1d		&	(id.)		&	\multicolumn{2}{c}{(id.)}	&	(id.)	&	(id.)	&	(id.)		& 		(id.)		& $\pm$2500		\\
Run 2{a,b,c}	& 	SKA1-mid$^b$	&	350 & 1760			&	1410	&	1	&	5		& $\approx11.6\times10^3$	& $\pm$270		\\
Run 3{a,b,c}	&	LOFAR-HBA$^c$	&	110 & 240			&	2600	&	0.05	& 	0.65		& $\approx$900			& $\pm$32		\\
\hline
\end{tabular}

\medskip
$^a$ Used in S16.
$^b$ Frequency coverage for SKA1-mid band 1+2 \protect\citep{dewdney2013}.
$^c$ \protect\cite{vanhaarlem2013}.
\end{table*}

\subsection{Setup of the Monte Carlo simulations}\label{MCsetup.sec}
For our Monte Carlo simulations we choose the three frequency setups listed in Table~\ref{mc_runs.tab}. 
The sources that we simulate emit at an intrinsic polarization angle $\chi_0 = 0$\degr, an RM of 0~\radm, and have a polarized flux density spectral index of -2, 0, or 2. 
All sources produce the same frequency-band-averaged polarized flux density, `$\langle L\rangle_{\Delta\nu}$', independent of their spectral index.
To ensure this the polarized flux density at the reference frequency of the source, $L_\mathrm{ref}$, satisfies
\begin{eqnarray}
\lefteqn{
\langle L\rangle_{\Delta\nu} = L_\mathrm{ref}\frac{\nu_\mathrm{ref}}{\Delta\nu}
} \nonumber\\
& &
\begin{cases}
\log{\left(\frac{\nu_2}{\nu_1}\right)} \hfill \alpha = -1\\
\frac{1}{\alpha+1}\left[\left(\frac{\nu_2}{\nu_\mathrm{ref}}\right)^{\alpha+1}-\left(\frac{\nu_1}{\nu_\mathrm{ref}}\right)^{\alpha+1}\right] \hfill \alpha \ne -1\, . 
\end{cases}
\end{eqnarray}
The observing band covers frequencies between $\nu_1$ and $\nu_2$, $\Delta\nu=\nu_2-\nu_1$, for $\nu_\mathrm{ref}$ we use the median frequency of the band.
All frequency channels are statistically independent and have Gaussian noise; the noise variances in Stokes $Q$ and $U$ are equal and constant.
If all channels have the same noise variance then combining $N_\mathrm{ch}$ frequency channels improves the signal-to-noise level of our observations by $\sqrt{N_\mathrm{ch}}$ (Section~\ref{errors.sec}). 
Therefore, to obtain signal-to-noise ratios in our Monte Carlo simulations of between zero and eight we set the standard deviation of the noise in each frequency channel equal to $\sqrt{N_\mathrm{ch}}$.

To each of the mock data sets that we generated this way we apply our method for finding the ML estimators of $Q_\mathrm{ref}$, $U_\mathrm{ref}$, RM, and $\alpha$.
The grid in (RM,$\alpha$) is searched out to RMs of $\pm$ 50 FWHM of the RM spread function and covers spectral indices between -6 and +2 (runs `a',`b', or `d') or between -12 and +12 (`c' runs); we also consider solutions to be valid if the ML estimators for RM or $\alpha$ fall within 1$\sigma$ of these boundaries.
The extent in $\alpha$ of the smaller grid reflects the range of spectral indices of known pulsars (\citealt{lorimer1995} and \citealt{bates2013}); this range in $\alpha$ is wide enough that it also covers the spectral indices of most known active galactic nuclei.
By using two different grid sizes we can test how important the choice for the size of the grid is; sources with a spectral index of +2 can only be investigated on the larger grid.
On the smaller grid runs 1a,b,d are based on 10$^4$ realisations of the Monte Carlo simulation. 
All other runs, including all runs on the larger grid, are based on 2000 realisations. 
Unless mentioned explicitly, names of runs refer to simulations on the smaller grid.

\begin{figure*}
\resizebox{\hsize}{!}{\includegraphics{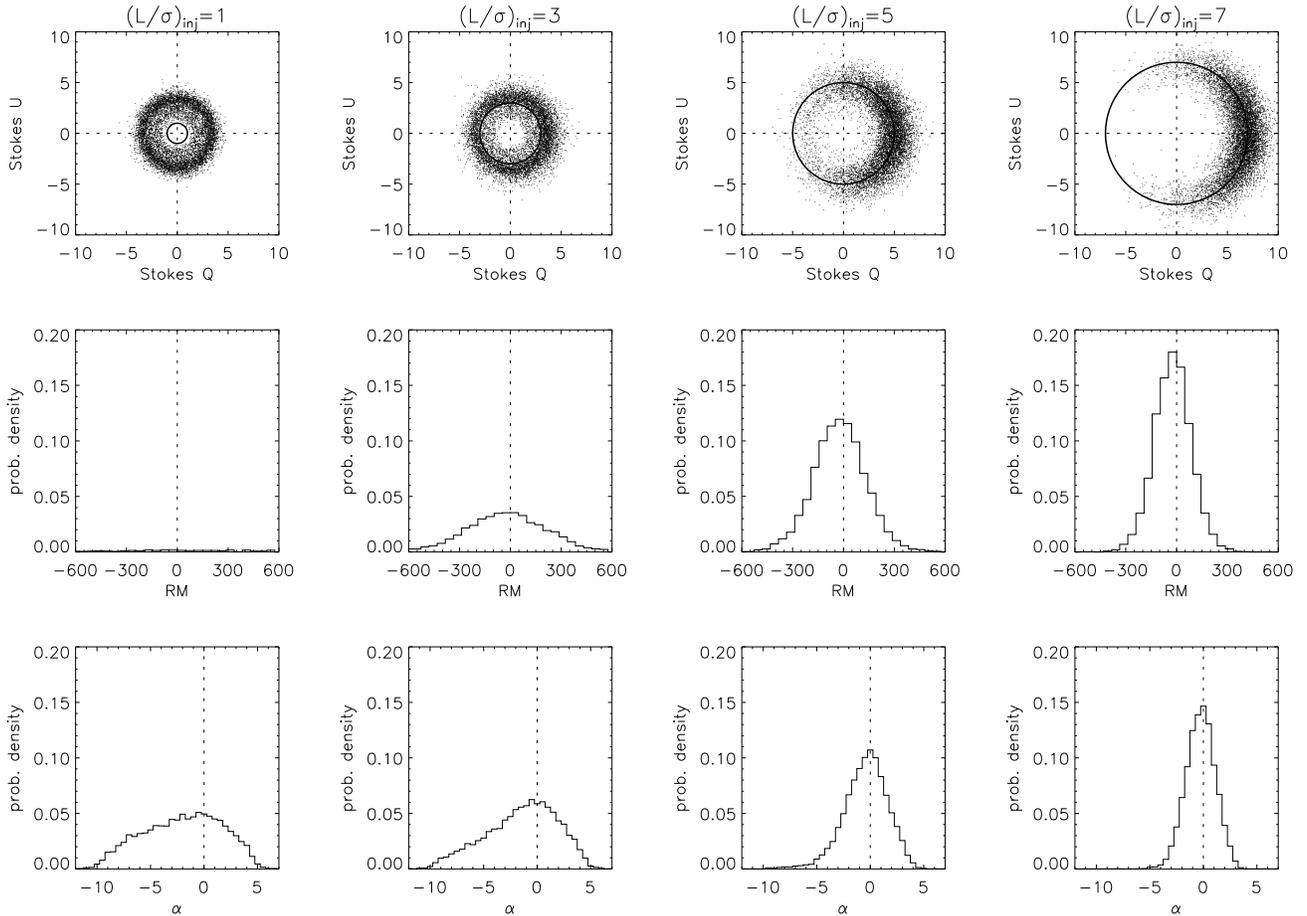}}
\caption{Distribution of the ML estimators for Stokes $Q_\mathrm{ref}$ and $U_\mathrm{ref}$, RM, and the spectral index $\alpha$, for simulation 1a and injected signal strengths between 1 and 7 sigma (after combining all frequency channels). The radius of the circle in the $Q,U$ diagrams is equal to the $L_\mathrm{ref}/\sigma$ ratio of the injected signal. For a signal-to-noise ratio of 1 the distribution of RM is very broad, extending well beyond the horizontal range of the panel, therefore the peak of this distribution is very low.
}
\label{overview.fig}
\end{figure*}

\subsection{A qualitative investigation of run 1a}
Fig.~\ref{overview.fig} shows the distributions of the ML estimators for $Q_\mathrm{ref}$ and $U_\mathrm{ref}$, RM, and $\alpha$ for run 1a for various strengths of the injected signal. 
At low signal-to-noise ratios noise often produces a higher peak in log likelihood in the (RM,$\alpha$) grid than the injected signal, and maximizing the log likelihood will then select the peak produced by noise (this also happens in RM synthesis: if the injected signal is weak compared to the noise level, selecting the highest polarized flux density in the RM spectrum misinterprets this peak as coming from the injected signal).
This leads to broad distributions in the ML estimators for $Q_\mathrm{ref}$, $U_\mathrm{ref}$, RM, and $\alpha$.
The larger the (RM,$\alpha$) grid, the higher the probability that noise will produce a higher log likelihood than the injected signal: we tested this with a simulation similar to run 1a except that we search $\alpha$ values up to $\pm$12.
The stronger the signal the more likely it is that the signal combined with noise produces the highest peak in log likelihood: this produces a transition between the `noise-dominated' and `signal-dominated' regimes. 
The latter occurs at an injected S/N of about 7 in Fig.~\ref{overview.fig}, where the distributions of RM and $\alpha$ become more Gaussian and the $Q_\mathrm{ref},U_\mathrm{ref}$ point cloud becomes distributed along the circle which indicates the signal-to-noise ratio of the injected signal.

Fig.~\ref{overview.fig} shows asymmetries, most clearly in the distribution of $\alpha$, which are related to the extent of the (RM,$\alpha$) grid.
At low signal levels the $\alpha$ distribution is very broad, and extends from the $\alpha = 0$ of the injected signal down to $\alpha$ values of less than -10, but up to $\alpha$ values of only +5.
This broad distribution is tapered at large positive or negative $\alpha$ values because we selected only solutions in our Monte Carlo simulations where the ML estimator of $\alpha$ lies between -6 and +2, or within 1$\sigma$ of this grid boundary.
A tail in the histogram of $\alpha$ values for weak injected signal persists even if we extend the search grid in $\alpha$ out to $\pm$12.
We found that the central hole in the $Q_\mathrm{ref},U_\mathrm{ref}$ distribution from Fig.~\ref{overview.fig} is not present in simulations 2a and 3a. 
By changing the fractional bandwidth of the simulations while keeping the centre frequency fixed we found that this gap appears in observations with small fractional bandwidths. 
The gap in the $Q_\mathrm{ref},U_\mathrm{ref}$ distribution is also filled in if we search out to $\alpha$ values of $\pm$12.

These results demonstrate that different choices for the frequency setup of observations or for the extent of the search grid in RM and $\alpha$ lead to different bias at low signal-to-noise ratios, in stark  contrast with the old situation where bias in only a single frequency channel was considered (e.g., fig.~2 in \citealt{wardle1974} and \citealt{simmons1985}).
Therefore, one should be cautious about generalizing results for polarization bias in wide-band data sets.

\subsection{Distributions of the polarized signal-to-noise ratio and of the detection significance}\label{signal-to-noise_distr.sec}
\begin{figure}
\resizebox{1.1\hsize}{!}{\includegraphics{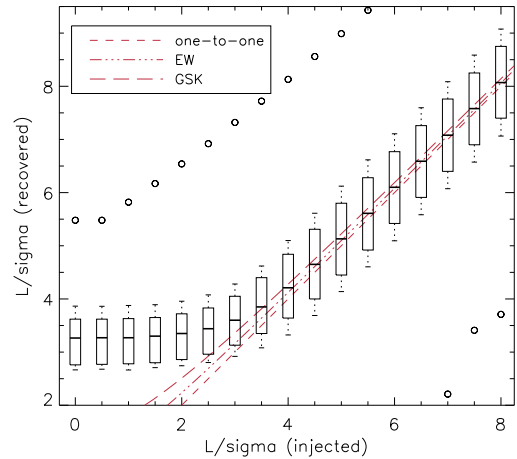}}
\caption{
Box-whisker plot comparing the injected polarized signal-to-noise ratio with the recovered ML estimators for the signal-to-noise ratio for run 1a. Each whisker extends out to 1.48 times the median absolute deviation, which, for Gaussian distributions, is $\approx 1\sigma$. Circles indicate the largest and smallest values encountered in the simulation.
The red lines show the one-to-one line and predictions for the debiasing methods selected by \citet[`EW']{everett2001} and by \citet[`GSK']{george2012}.
}
\label{snr_comparison.fig}
\end{figure}

Fig.~\ref{snr_comparison.fig} compares the injected to the recovered ML estimators for the polarized flux density divided by the standard deviation of the noise; also shown are predictions for the debiasing methods selected by \cite{everett2001} and \cite{george2012}
(note that Everett et al. proposed their method for debiasing individual frequency channels, while we considered the entire frequency band).
The bias correction selected by Everett et al. is based on work by \cite{wardle1974} and \cite{simmons1985}.
We use box-whisker plots because these provide information also for asymmetric distributions and/or distributions with outliers.
At high signal-to-noise levels the vertical dotted lines in each box-whisker symbol extend to what would be $\pm1\sigma$ in a 1D Gaussian distribution.

Fig.~\ref{snr_comparison.fig} shows that weak signals (signal-to-noise ratio $\lesssim$ three) are often reconstructed at a similar signal-to-noise ratio of about three independent of the strength of the injected signal (\citealt{hales2012} reached a similar conclusion in their analysis, see fig.~1 in their paper).
Because the measured signal-to-noise ratio can be interpreted in this case by a wide range of injected signal-to-noise ratios, debiasing makes little sense.
The height of the plateau at low signal-to-noise levels depends on the frequency setup of the simulation, as we found by comparing the equivalents of Fig.~\ref{snr_comparison.fig} between runs 1a, 2a, and 3a: run 2a shows the weakest noise bias (median recovered signal-to-noise ratio of about two).

In run 1a, at signal-to-noise levels $\gtrsim$ five the median of the distribution of reconstructed signal-to-noise ratios deviates little from the injected signal strength, indicating negligible bias.
At lower signal-to-noise levels these medians follow the debiasing curve proposed by Everett et al. more closely than the debiasing curve proposed by George et al., even though the scatter in the recovered signal-to-noise values is much larger than the separation between the debiasing curves.  As we noted at the end of the previous subsection, different frequency setups are affected by noise in different ways. Since George et al. proposed their debiasing method for observations centred on 1.4 GHz, using a narrow bandwidth, their debiasing method might not work as well for our simulation run 1a.
If the source has a spectral index of -2 then the recovered signal-to-noise ratios do not increase as rapidly as the injected signal-to-noise ratios if the signal-to-noise ratio is high; the equivalent of fig.~\ref{snr_comparison.fig} for run 2a even shows a local minimum. 
This bias is reduced strongly if the spectral index is +2 (runs 1-3).

\begin{figure}
\resizebox{\hsize}{!}{\includegraphics{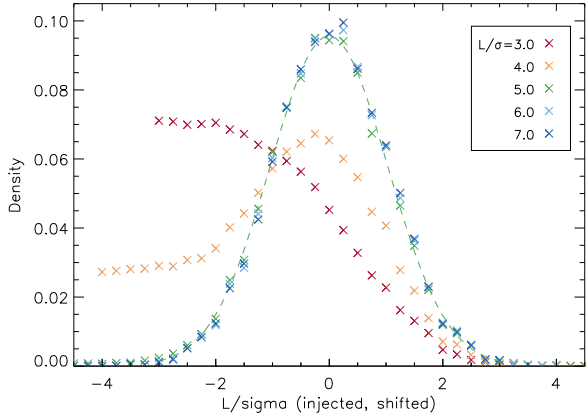}}
\caption{
Translation from \emph{measured} signal-to-noise ratios between 3 and 7 (indicated by different colours) back to injected signal-to-noise ratios for simulation 1a.
These distributions correspond to horizontal cuts in Fig.~\ref{snr_comparison.fig} after gridding this figure using cells with a width of 0.5 units.
The pdfs were centred on the value of the measured $L/\sigma$ of each cut, i.e., this figure shows $\mathrm{pdf}\left[(L/\sigma)_\mathrm{injected}\right] - (L/\sigma)_\mathrm{observed}$.
We indicated this by using `shifted' along the x-axis.
For measured signal-to-noise ratios larger than about five the distribution of injected signal-to-noise values can be fitted well by a Gaussian, as indicated by the dashed line.
The fitted Gaussian which we plotted has a standard deviation of 1.03.
}
\label{snr_intrinsic.fig}
\end{figure}

An important question is how the measured signal-to-noise ratio can be translated back into the signal-to-noise ratio of the injected signal.
This is equivalent to drawing a horizontal line in Fig.~\ref{snr_comparison.fig} and determining the distribution of injected $L/\sigma$ along this line.
To solve this inverse problem we divided Fig.~\ref{snr_comparison.fig} into a fine grid (cell size 0.5$\times$0.5) and counted the number of realisations of the Monte Carlo simulation in each cell. 
Then we read off along a horizontal line (i.e. for a given observed $L/\sigma$) which injected $L/\sigma$ values contribute. 
Note that this method only works if the number of realisations of the Monte Carlo process is the same for each injected $L/\sigma$.
Fig.~\ref{snr_intrinsic.fig} shows for five different strengths of the measured signal which strengths of the injected signal contribute.
For weak signals a Gaussian distribution is definitely not a good fit.
If the measured signal-to-noise ratio is five then the distribution of injected $L/\sigma$ values is described well by a Gaussian with a standard deviation of almost one (dashed green line); if the measured signal-to-noise ratio is even higher then this standard deviation is even closer to one.
Also, if the measured $L/\sigma$ is five or larger then the fitted Gaussians have a mean close to zero, which implies that in these cases the distribution of the injected $L/\sigma$ values is centred on the value of $L/\sigma$ of the measured signal. 
Therefore, for these signal-to-noise ratios the distributions of $L/\sigma$ are almost bias-free\footnote{For measured $L/\sigma$ of 5,6,7 the distributions of injected $L/\sigma$ shown in Fig.~\ref{snr_intrinsic.fig} have means of -0.01,0.03,0.03.
To understand whether these means are consistent with zero (indicating no bias), we applied the one-sample $t$-test. 
We assume that the distributions of injected $L/\sigma$ values are close enough to being Gaussian that we can run this test.
The $t$-values for the three distributions are -0.58, 2.50, and 3.41; larger values for $|t|$ are encountered in 56\%, 1.2\%, and 0.07\% of cases, respectively, hinting at bias for the highest signal-to-noise levels.
}. 
We could not extend our analysis to even larger values for the measured signal-to-noise ratio, because for such strong signals the distributions of injected $L/\sigma$ extend beyond the grid which we used to make Fig.~\ref{snr_intrinsic.fig}.

\begin{figure}
\resizebox{\hsize}{!}{\includegraphics{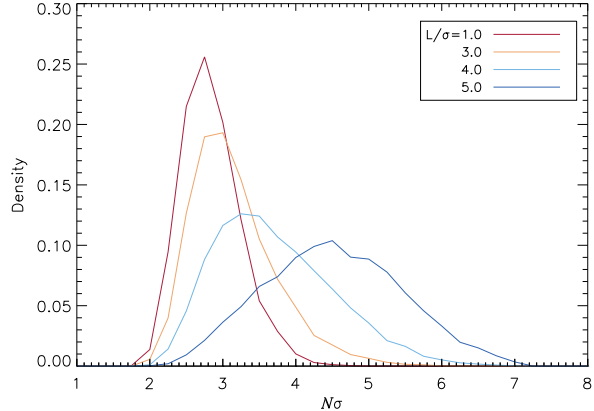}}
\caption{
Distribution of detection significances which we calculated from the log likelihood ratio for four strengths of the injected signal (simulation 1a, using a binsize of 0.2 along the horizontal axis). 
To make the detection significances easier to interpret, we calculate for each detection how far we have to integrate the error function to obtain the same detection significance (`$N\sigma$', which is plotted along the horizontal axis). 
The distributions for signal-to-noise ratios of 0 and 1 are nearly identical; to avoid cluttering the plot we omit the distribution for a signal-to-noise ratio of 2. 
}
\label{prob_hist.fig}
\end{figure}

Closely related to the distribution of the ML estimators for the polarized signal-to-noise ratio is the distribution of the detection significance. The latter is quantified by the difference in log likelihood for a realisation with and without the injected signal, which can be thought of as the height of the peak in the log likelihood landscape compared to its surroundings. A stronger signal (higher signal-to-noise ratio) will produce a higher peak in log likelihood, leading to a higher detection significance. 
As Fig.~\ref{snr_comparison.fig} shows, this correlation between the detection significance and the signal-to-noise ratio of the measured signal breaks down if the injected signal is weak ($L/\sigma \lesssim 3$) and the noise contribution is responsible for producing the highest peak in the log likelihood landscape.
Because we are more familiar with quantifying detection significances using the error function, we calculate for each detection the value of $N\sigma$ for which the error function gives the same detection significance as the significance we derive from the log likelihood ratio.
Fig.~\ref{prob_hist.fig} shows the distributions of $N\sigma$ for a range of strengths of the injected signal in simulation 1a.
If the injected signal is weak then noise can produce the highest log likelihood in a large (RM,$\alpha$) grid, similar to how a large map of the sky can show bright spots due to noise.
This leads to many realisations in the Monte Carlo process having a detection significance above `$3\sigma$' even if the injected signal has only $L/\sigma$=1.
A sufficiently high detection threshold, e.g., `5$\sigma$', will prevent such weak sources being interpreted as real detections, but at the same time Fig.~\ref{prob_hist.fig} shows that some genuine signals will fall below this detection threshold and will therefore be discarded (false negatives).
Fig.~\ref{prob_hist.fig} also shows that for $L/\sigma$=4,5 the `$N\sigma$' coordinate of the mean of each distribution is smaller than the value of $L/\sigma$ of the injected signal.
This implies that quoting the measured $L/\sigma$ of a detection overestimates the real detection significance, because the centres of the distributions of injected and measured $L/\sigma$ in Fig.~\ref{snr_intrinsic.fig} are almost the same for these two $L/\sigma$ values.

\subsection{Distribution of RMs}\label{RM_distr.sec}
\begin{figure}
\resizebox{\hsize}{!}{\includegraphics{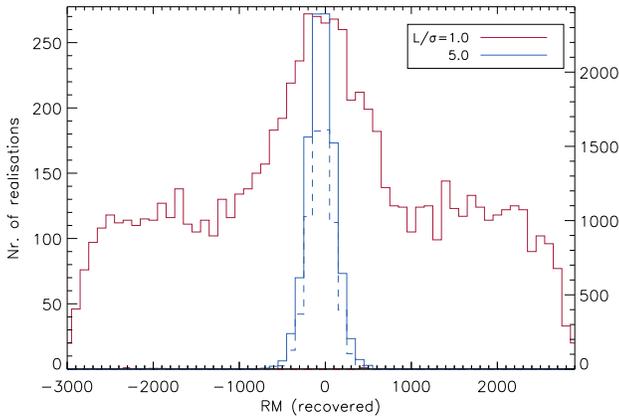}}
\caption{Derived RMs for simulation 1d for injected signal-to-noise strengths of 1 (red, left y axis) and 5 (blue, right y axis). Dashed lines include only data points with an overall detection significance $>\, `4\sigma$' (see main text). 
}
\label{rm_hist.fig}
\end{figure}

Weak signals produce broad RM distributions, as demonstrated in Fig.~\ref{overview.fig}. 
If we limit the search range in RM (run 1d) we can make the effects of noise on the RM distribution for different strengths of the injected signal even more clear, as shown in Fig.~\ref{rm_hist.fig} (see also fig.~3 in \citealt{george2012} and fig.~2 in \citealt{macquart2012}).
These broad distributions can be removed by selecting only realisations of the Monte Carlo simulations which have detection probabilities larger than `$4\sigma$'. 
As Fig.~\ref{rm_hist.fig} shows this removes almost all the realisations belonging to the injected signal with $L/\sigma=1$, but also if $L/\sigma=5$ a large fraction of the realisations is removed.
Why this happens can be understood from Fig.~\ref{prob_hist.fig}, which shows that even if the injected signal has $L/\sigma = 5$ a large proportion of the Monte Carlo realisations has a detection significance smaller than `$4\sigma$'.

\begin{figure}
\resizebox{\hsize}{!}{\includegraphics{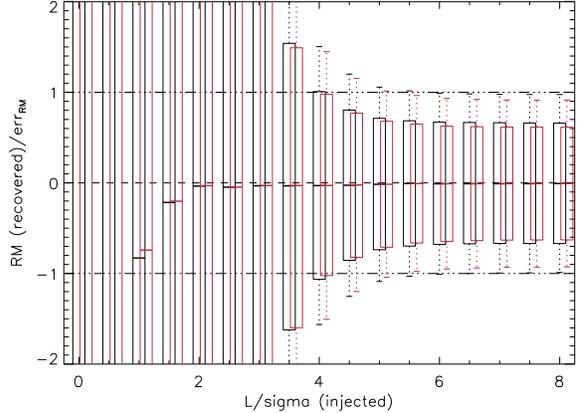}}
\caption{Box-whisker plot showing the distribution of reconstructed $\widehat{\mathrm{RM}}$ divided by either the measurement error as derived from fitting a 2D parabola to contours in the $\log{\Lambda}$ landscape (black lines) or by the standard error in RM, err$_\mathrm{RM}$ (red lines; points are slightly offset along the horizontal axis).
Simulations for run 1a; note that at low S/N the error estimation in RM using log likelihood fitting is unreliable (Section~\ref{errors.sec}).
}
\label{rm_errors.fig}
\end{figure}

Often the error in the measured RM is derived from $\mathrm{err}_\mathrm{RM} = \mathrm{FWHM}/\left(2\, L/\sigma\right)$, where $\mathrm{FWHM}$ is the full width at half-maximum of the RM spread function (which can be calculated using equation~61 in \citealt{brentjens2005}), $L$ the measured polarized flux density, and $\sigma$ the rms noise level of the observations when integrating over the entire frequency band\footnote{
To derive this expression start with equation~A.18 in \citealt{brentjens2005}. If all channels have the same noise variance and the source has a spectral index of zero, and assuming a uniform coverage in wavelength squared instead of frequency which is sampled with $N_\mathrm{ch}$ channels, one can show that $\sigma_{\lambda^2}^2 = N_\mathrm{ch}(N_\mathrm{ch}+1)(\delta\lambda^2)^2/12$, with $\delta\lambda^2$ the width of each channel. Since $\Delta\lambda^2=N_\mathrm{ch}\delta\lambda^2$, $\sigma_\mathrm{RM}= \sqrt{N_\mathrm{ch}/\left[(N_\mathrm{ch}-2)(N_\mathrm{ch}+1)\right]}\left(\sigma/L\right)\left(\sqrt{3}/\Delta\lambda^2\right)$. For $N_\mathrm{ch}\gg 1$ the square root $\approx 1/\sqrt{N_\mathrm{ch}}$, while $\sqrt{3}/\Delta\lambda^2$ is half the FWHM of the RM spread function. Noting that the signal-to-noise ratio after RM synthesis increases as $\sqrt{N_\mathrm{ch}}$ (see, e.g., Section~\ref{errors.sec}), one then recovers the standard expression for the measurement error in RM after running RM synthesis.}.
In Fig.~\ref{rm_errors.fig} we test if the distribution of RM/err$_\mathrm{RM}$ derived from observations is biased, and if $\mathrm{err}_\mathrm{RM}$ is a good approximation for the error in RM.
For large enough injected $L/\sigma$ values the distribution of recovered RM/err$_\mathrm{RM}$ is centred on the injected RM of 0~\radm, implying that this estimator is unbiased.
As indicated by the black whiskers, the width of this distribution is (almost) one if we use the errors calculated from the log likelihood, as one would expect if the calculated errors are correct.
However, the width of the distribution of RM/err$_\mathrm{RM}$ is noticeably different from one if $\mathrm{err}_\mathrm{RM}$ is used. 
So far we have not found an explanation why err$_\mathrm{RM}$ is only approximately correct; not debiasing $L$ before calculating err$_\mathrm{RM}$ can be ruled out, because Fig.~\ref{snr_comparison.fig} shows that noise bias in $L$ is negligible for $L/\sigma \gtrsim 5$.

There are only small differences in the extent of the `$1\sigma$' confidence regions in Fig.~\ref{rm_errors.fig} between runs 1a, 2a, and 3a if we estimate measurement errors in RM by fitting contours in the log likelihood landscape; using err$_\mathrm{RM}$ instead leads to much larger differences.
In fact, our Monte Carlo simulations show that using err$_\mathrm{RM}$ can lead to `$1\sigma$' whiskers which are much smaller or larger than $\pm1$ in Fig.~\ref{rm_errors.fig}.
If the source has a spectral index of -2 then the `$1\sigma$' whiskers in Fig.~\ref{rm_errors.fig} extend only out to about $\pm 0.3$ in run 2b, and out to about $\pm 0.75$ in run 3b, if we use err$_\mathrm{RM}$.
Two effects play a role.
First, a source with a negative spectral index has a smaller error in RM than a source with a positive spectral source if the same frequency setup is used in both cases: a negative spectral index implies that the source is bright at low frequencies (large $\lambda^2$) where RMs can be determined accurately, while the opposite is true for a source with a positive spectral index.
Since err$_\mathrm{RM}$ is calculated assuming $\alpha=0$, this suggests that sources with negative (positive) $\alpha$ on average have smaller (larger) errors than the value calculated with err$_\mathrm{RM}$. When dividing the ML estimator for RM by err$_\mathrm{RM}$ this leads to confidence regions which extend to less than (more than) $\pm 1$ along the vertical axis in Fig.~\ref{rm_errors.fig}.
Second, err$_\mathrm{RM}$ depends on the signal-to-noise ratio of the signal, and these can be strongly biased as we reported in Section~\ref{signal-to-noise_distr.sec}.
By replacing the measured signal-to-noise ratio in err$_\mathrm{RM}$ with the injected signal-to-noise ratio we found that this second effect by itself is not sufficient to explain the behaviour of the distributions of RM/err$_\mathrm{RM}$.

By comparing the equivalents of Fig.~\ref{rm_errors.fig} for different simulation runs we find that the rate at which the error in RM decreases with increasing signal-to-noise level depends both on the frequency setup of the observations and on the spectral index of the source.
For example, if $\alpha=2$ and we use err$_\mathrm{RM}$ then the `$1\sigma$' whiskers in Fig.~\ref{rm_errors.fig} extend well beyond $y=\pm1$, falling outside the plot range (run 2c) or extending out to $y=\pm1.4$ (run 3c).
If we determine the error in RM by fitting contours in log likelihood then in the case of run 2c the `$1\sigma$' whiskers decrease towards $y=\pm1.3$ at a signal-to-noise ratio of eight; we did not investigate if the whiskers converge to $y=\pm1$ at even higher signal-to-noise ratios.

We conclude that err$_\mathrm{RM}$ provides a fast way for estimating the error in RM for a given frequency setup and observing time, but more accurate methods are required for determining measurement errors in RM for sources with non-zero spectral indices.

\begin{figure}
\resizebox{\hsize}{!}{\includegraphics{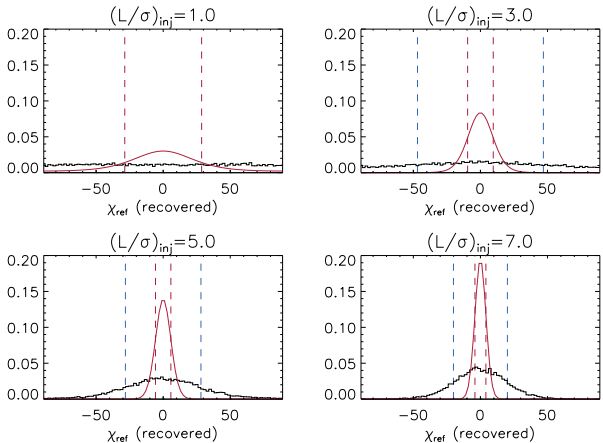}}
\caption{Probability density functions of the intrinsic polarization angle $\chi_\mathrm{ref}$ (black lines) for four different strenghts of the injected signal (run 1a). The solid red line shows the distribution from equation~3 in \protect\cite{naghizadeh-Khouei1993}, the vertical red lines show the estimate for the measurement error $\left[1/\left(2 \left(L/\sigma\right)_\mathrm{ch}\right)\right]/\sqrt{N_\mathrm{ch}}$, and the vertical blue lines show the prediction by equation~\ref{err_chi.eqn}.
}
\label{chidistr.fig}
\end{figure}

\subsection{Distributions of the intrinsic polarization angle $\chi_\mathrm{ref}$ and the spectral index $\alpha$}\label{PA_distr.sec}
\cite{naghizadeh-Khouei1993} discussed the probability density function for the measured polarization angle of a single frequency channel. These authors show that at high signal-to-noise levels this distribution has a standard deviation of $1/\left(2 \left(L/\sigma\right)_\mathrm{ch}\right)$ radians, where $\left(L/\sigma\right)_\mathrm{ch}$ is the signal-to-noise ratio of a single frequency channel.
However, the intrinsic polarization angle $\chi_\mathrm{ref}$, which we derive from the ML estimators for $Q_\mathrm{ref}$ and $U_\mathrm{ref}$, has been corrected for Faraday rotation; therefore the distribution of $\chi_\mathrm{ref}$ from our Monte Carlo simulations will differ from the probability density function derived by Naghizadeh-Khouei and Clarke (Fig.~\ref{chidistr.fig}).
To estimate the measurement uncertainty ($1\sigma$) in $\chi_\mathrm{ref}$ we rewrite equation~A.20 from \cite{brentjens2005}, assuming uniformly spaced channels in wavelength squared, constant noise variances across the band (equal in Stokes $Q$ and $U$), $\delta\lambda^2 \ll 1$, and $N_\mathrm{ch} \gg 1$: 
\begin{eqnarray}
\mathrm{err}_\chi = \frac{1}{2\left(L/\sigma\right)_\mathrm{ch}\sqrt{N_\mathrm{ch}}}\sqrt{1+3\left(\frac{\lambda^2_\mathrm{max}+\lambda^2_\mathrm{min}}{\lambda^2_\mathrm{max}-\lambda^2_\mathrm{min}}\right)^2}\, ,
\label{err_chi.eqn}
\end{eqnarray}
where $\lambda^2_\mathrm{min}$ and $\lambda^2_\mathrm{max}$ are the lowest respectively the highest wavelength squared values in the observing band. 
One can interpret this equation as the product of the error in $\chi_\mathrm{ref}$ derived by Naghizadeh-Khouei and Clarke after summing $N_\mathrm{ch}$ channels, modulated by a term which contains information on the frequency coverage of the observations (the ratio inside the square root is proportional to one over the fractional bandwidth expressed in units of wavelength squared).

In simulations 1-3a,b,c the means of the distributions of $\chi_\mathrm{ref}$ lie between $\pm2.2$\degr. 
Such small offsets imply that the bias in $\chi_\mathrm{ref}$ is small; in fact, these offsets could be largely due to the limited number of realisations of the Monte Carlo process.
Using the signal-to-noise ratio of the injected signal in equation~\ref{err_chi.eqn}, the standard deviations of the $\chi_\mathrm{ref}$ distributions shown in Fig.~\ref{chidistr.fig} can be described well by equation~\ref{err_chi.eqn}, with the exception of the first panel where the signal-to-noise level is simply too low.
In simulations 1-3a,b,c we found that injected signal-to-noise ratios of 5 and 7 can have differences between the prediction by equation~\ref{err_chi.eqn} and the standard deviation of the $\chi_\mathrm{ref}$ distributions of up to four degrees; for weaker signals these differences can be (much) larger.
Based on these observations we conclude that equation~\ref{err_chi.eqn} is a useful approximation for estimating the error in the derived polarization angle if the signal is sufficiently strong.

As demonstrated first by \cite{jauncey1967}, maximizing the log likelihood for the coefficient $\alpha$ in a power-law distribution typically requires numerical techniques. The expression we derived for 
finding the ML estimator of $\alpha$ in the case of equal noise variances,
\begin{eqnarray}
\lefteqn{\sum_{i=1}^{N_\mathrm{ch}}
\left[
Q_{\mathrm{obs},i}\left(\hat{Q}_\mathrm{ref}\,c_i - \hat{U}_\mathrm{ref}\,s_i\right) \right. 
} \nonumber \\
 & + & \left.
U_{\mathrm{obs},i}\left(\hat{Q}_\mathrm{ref}\,s_i + \hat{U}_\mathrm{ref}\,c_i\right) \right]
\left(\frac{\nu_i}{\nu_\mathrm{ref}}\right)^\alpha \log\left(\frac{\nu_i}{\nu_\mathrm{ref}}\right)/\sigma_{L,i}^2
 \nonumber \\
 & = & 
\sum_{i=1}^{N_\mathrm{ch}} (\hat{Q}_\mathrm{ref}^2 + \hat{U}_\mathrm{ref}^2) 
\left(\frac{\nu_i}{\nu_\mathrm{ref}}\right)^{2\alpha} \log\left(\frac{\nu_i}{\nu_\mathrm{ref}}\right)/\sigma_{L,i}^2 \, ,
\end{eqnarray}
also cannot be solved analytically. Therefore we did not investigate further if analytical expressions can be found for the probability density function of $\alpha$ or the standard deviation of this distribution.

\subsection{Summary of main results from Section 3}\label{summary.sec}
In this Section we investigated how source parameters determined from noisy data are affected by the frequency setup of the observations, the spectral index of the source, and the extent of the search grid in RM and $\alpha$.
All three factors influence the results, which implies that debiasing schemes should include all these factors and that such schemes are difficult to generalize.
Debiasing weak signals makes little sense, since the injected signal one wishes to recover from the observations gets lost in the noise.
Citing the polarized signal-to-noise ratio of a measurement is a biased proxy for the detection significance; the log likelihood ratio is much more accurate.
The measured signal-to-noise ratio is itself subject to noise, and can be explained by a range of possible injected signal-to-noise ratios.
For sufficiently strong signals the latter distribution is Gaussian.
We derive the equation for the error in RM that is often used in the literature, $\mathrm{err}_\mathrm{RM} = \mathrm{FWHM}/(2 L/\sigma)$, and show that this quantity systematically under- or overestimates the true measurement uncertainty in RM. 
Sources with spectral indices $\alpha \ne 0$ typically have measurement uncertainties in RM which are different from $\mathrm{err}_\mathrm{RM}$.
Finally, we derive an expression for the measurement uncertainty in the intrinsic polarization angle, $\mathrm{err}_\chi$. 
Since this derivation is based on the same assumptions as our derivation of $\mathrm{err}_\mathrm{RM}$, also $\mathrm{err}_\chi$ should be considered an approximation which can deviate from the true measurment error.

\section{Conclusions}\label{conclusions.sec}
In this paper we developed two methods for determining the physical properties of radio sources which emit at one RM, i.e., by assuming a power-law polarized flux density spectrum or that the polarized flux density spectrum is a scaled version of the spectrum in Stokes $I$. We allow for a variation in the sensitivity (noise variance) across the observing band. By fitting contours in the log likelihood landscape we determine the ML estimators for the source parameters, the measurement uncertainties, and the detection significance. Also we derive an expression to quickly estimate the measurement error in the intrinsic polarization angle. 

If the polarized flux density follows a power law with a spectral index $\alpha$, then maximizing the likelihood requires numerically searching through a grid of (RM,$\alpha$) values; the ML estimators for the Stokes $Q$ and $U$ flux densities at the reference frequency ($Q_\mathrm{ref}$ and $U_\mathrm{ref}$) and the multiplicative scale factor for the noise variance $\eta$ can be calculated analytically for each grid point. In this case we show that for a source with $\alpha=0$ the ML estimators for $Q_\mathrm{ref}$ and $U_\mathrm{ref}$ can be found by applying RM synthesis. 
We also show that a variation in the sensitivity across the observing band can be included in the RM synthesis formalism by using one over the noise variance in each frequency channel as weights.

If the polarized flux density spectrum is a scaled version of the Stokes $I$ spectrum then the ML estimators for RM and the polarization fractions $q$ and $u$ can only be found using numerical techniques. 
Applying RM synthesis to the observed flux density ratios $Q_{\mathrm{obs},i}/I_{\mathrm{obs},i}$ and $U_{\mathrm{obs},i}/I_{\mathrm{obs},i}$ (where `$i$' is the channel index) does not maximize the likelihood, because the equations for finding the ML estimators for $q$ and $u$ do not depend on these ratios of flux densities.
However, for weakly polarized sources we derive a weighted form of RM synthesis which includes the Stokes $I$ spectrum of the source and a variation in sensitivity across the frequency band.

For sources with a power-law flux polarized density spectrum we use Monte Carlo simulations to investigate statistical bias and whether standard methods for estimating measurement uncertainties are accurate. 
We simulate different frequency setups (4 cm band on the ATCA, bands 1+2 on SKA1-mid, and LOFAR HBA), sources with spectral indices $\alpha$ of -2, 0, or +2, and different extents of the search grid in RM and $\alpha$.  

We find that noise bias affects different frequency bands in different ways, which means that in the low signal-to-noise regime results for one frequency band cannot be generalized.
For observations in the 4 cm band of the ATCA noise bias in the polarized flux density is negligible if the signal-to-noise ratio is larger than about five. 
An observed signal-to-noise ratio can be the result of a range of injected signal-to-noise ratios plus noise. 
We found that this distribution of injected signal-to-noise ratios is approximately Gaussian if the measured signal-to-noise ratio is larger than five.
We also found that citing the polarized signal-to-noise ratio as a proxy for the detection significance overestimates this significance.

At low signal-to-noise ratios the distribution of the ML estimators for RM and the spectral index is very wide.
For weak signals the highest peak in the likelihood can be produced purely by noise, or a combination of noise plus the injected signal.
In these situations the ML estimators for the model parameters can be very different from the properties of the injected (true) signal.
Also RM synthesis is susceptible to this, because in that case one interprets the highest peak in polarized flux density in the RM spectrum as being due to a real signal.
The standard method for calculating the measurement error in RM only applies to sources with $\alpha=0$; in certain cases the combination of noise bias in the polarized flux density and spectral index effects leads to large differences between the true measurement error in RM and the error which is calculated using the standard method. 
Finally, we derive an equation for approximating the error in the intrinsic polarization angle, $\mathrm{err}_\chi$, and show that this quantity depends both on the signal-to-noise ratio and on the frequency coverage of the observations.

\section*{Acknowledgements}
We would like to thank Aristeidis Noutsos and Olaf Wucknitz (both at the Max Planck Institute for Radio Astronomy) for many fruitful discussions.
We also thank the anonymous referee for their very constructive comments.
KJL is supported by the National Basic Research Program of China, 973 Program (Grant no. 2015CB857101) and NSFC (Grant no. U15311243).
The figures shown in our paper make use of colour tables that were developed by Paul Tol\footnote{https://personal.sron.nl/$\sim$pault/} (Netherlands Institute for Space Research).

\bibliography{/Users/sch507/articles/ne6e} 

\appendix
\section{Nomenclature}\label{nomenclature.sec}
In the scientific literature both `RM' and `Faraday depth' are used for the integral in equation~\ref{rm_definition.eqn}, having been introduced by \cite{gardner1963} and \cite{burn1966}, respectively.
In this Appendix we explain why we prefer to use `RM'. 
Compared to our argument in \cite{schnitzeler2015b} we now allow for synchrotron emission and Faraday rotation to take place in the same part of the line of sight d$l$.

It is not clear why \cite{burn1966} did not consider keeping the term `RM'. 
Considering the radiative transfer of polarized radio waves, as presented by e.g., \cite{sazonov1969}, \cite{jones1977}, and more recently by \cite{huang2011}, it is sensible to use `RM' instead of `Faraday depth'.
Equation~1 of \cite{jones1977} showed that, in a non-relativistic, optically thin source (i.e. with absorption neglected) where 
Faraday conversion between linear and circular polarization is negligible (this is the medium considered by Burn in his paper) the radiative transfer of the linear polarization vector can be written as
\begin{eqnarray}
\frac{\mathrm{d}\bmath{L}_\nu}{\mathrm{d}l}\, =\, 2\mathrm{i}\times0.81n_\mathrm{e}B_\|\bmath{L}_\nu + \epsilon_\nu\mathrm{e}^{2\mathrm{i}\chi_0}\, ,
\label{transfer_L.eqn}
\end{eqnarray}
where $\epsilon_\nu$ is the volume emissivity of the source.
All parameters in equation~\ref{transfer_L.eqn} can vary along the line of sight.
The physical interpretation of equation~\ref{transfer_L.eqn} is that each infinitesimal path element d$l$ rotates background synchrotron emission and adds its own locally generated emission. 
From the perspective of the polarization vector, from the moment it is emitted it will only encounter Faraday rotating screens when travelling in the direction of the observer. 
Before Burn wrote his paper, the tiny amount of Faraday rotation that each path length d$l$ adds, $0.81n_\mathrm{e}B_\| \mathrm{d}l$ in equation~\ref{transfer_L.eqn}, would have been known as the RM of the infinitesimal path length d$l$. 
Because radiative transfer can be expressed completely in terms of RM, Burn could have kept using this term and did not need to introduce the concept of Faraday depth.

Of course, one cannot use `RM' both for the integral in equation~\ref{rm_definition.eqn} and to indicate the change in polarization angle with increasing wavelength squared; the numerical value of this derivative is only equal to RM if the source emits at a single RM.
We used the term `net RM' in \cite{schnitzeler2015b} to indicate the change in polarization angle with wavelength squared: since the observed monochromatic polarization vector is the beam-averaged sum of all the polarization vectors emitted by the source, both the angle of this monochromatic polarization vector and also the change in angle with wavelength squared are net quantities which depend on the contributions by all infinitesimal source elements.

\section{Finding ML estimators for $\lowercase{q}$ and $\lowercase{u}$ if $L \propto I$}\label{Appendix_A}
In this appendix we derive the equations for finding the ML estimators of $q$ and $u$ if the polarized flux density spectrum of a source is a scaled version of its Stokes $I$ spectrum. 
There are two ways for removing the nuisance parameters  $I_\mathrm{mod,i}$, the modelled Stokes $I$ flux density in each channel: either one finds the ML estimators for $I_\mathrm{mod,i}$ (Section~\ref{exact_noise_variances.sec}), or one marginalizes over these parameters (Section~\ref{marginalization.sec}). Because each option leads to different equations for $\hat{q}$ and $\hat{u}$, we consider them in separate sections.

\subsection{Solve for $I_{\mathrm{mod},i}$}\label{exact_noise_variances.sec}
We derive the ML estimators for $\eta$ and $I_\mathrm{mod,i}$ by taking the partial derivatives of the log likelihood with respect to these parameters and setting the result equal to zero.
$\hat{\eta}$ depends on $q, u$, and $I_\mathrm{mod,i}$, while $\hat{I}_\mathrm{mod,i}$ depends on $\hat{\eta}$ and therefore recursively on $I_\mathrm{mod,i}$. 
Solving for $I_\mathrm{mod,i}$ and subsequently finding the ML estimators for $q$ and $u$ is difficult.

To investigate whether it is at all possible to find analytical solutions for the ML estimators of $q$ and $u$ we make the simplifying assumptions that 1) the measured noise variances are exact ($\eta$ = 1) and 2) $\sigma_{Q,i}=\sigma_{U,i}=\sigma_{I,i}=\mathrm{constant}$. 
In this case it is easy to find the ML estimator for $I_\mathrm{mod,i}$.
The partial derivatives of the log likelihood with respect to $q$ and $u$ have in their denominators $\sigma^2\left(q^2+u^2+1\right)^2$  and in their numerators
\begin{eqnarray}
\lefteqn{\left(-1+q^2-u^2\right)\left(\mathcal{QI}+u\,\mathcal{QU}\right) - qu^2\left(\mathcal{QQ}-\mathcal{UU}\right) } \nonumber\\
 & & + 2qu\,\mathcal{UI} + q\left(\mathcal{II}-\mathcal{QQ}\right)\, \label{dlogLdq_nomargin.eqn} \\	
\lefteqn{\left(1+q^2-u^2\right)\left(\mathcal{UI}+q\,\mathcal{QU}\right) - q^2u\left(\mathcal{QQ}-\mathcal{UU}\right) } \nonumber\\
 & & - 2qu\,\mathcal{QI} - u\left(\mathcal{II}-\mathcal{UU}\right)\, ,
\label{dlogLdu_nomargin.eqn}
\end{eqnarray}
where we used the following shorthand notation for the observables:
\begin{eqnarray}
Q_{\mathrm{derot},i} & = & c_i Q_{\mathrm{obs},i} + s_i U_{\mathrm{obs},i}  \\
U_{\mathrm{derot},i} & = & -s_i Q_{\mathrm{obs},i} + c_i U_{\mathrm{obs},i}  \\
\mathcal{QQ} & = & \sum_{i=1}^{N_\mathrm{ch}} Q_{\mathrm{derot},i}^2  \\
\mathcal{UU} & = & \sum_{i=1}^{N_\mathrm{ch}} U_{\mathrm{derot},i}^2  \\
\mathcal{II} & = & \sum_{i=1}^{N_\mathrm{ch}} I_{\mathrm{obs},i}^2  \\
\mathcal{QI} & = & \sum_{i=1}^{N_\mathrm{ch}} Q_{\mathrm{derot},i} I_{\mathrm{obs},i}  \\
\mathcal{QU} & = & \sum_{i=1}^{N_\mathrm{ch}} Q_{\mathrm{derot},i}U_{\mathrm{derot},i}  \\
\mathcal{UI} & = & \sum_{i=1}^{N_\mathrm{ch}} U_{\mathrm{derot},i} I_{\mathrm{obs},i}\, . 
\end{eqnarray}
To maximize the likelihood it is sufficient to find the zero-points of the numerators of $\partial \log \Lambda /\partial q$ and $\partial \log \Lambda /\partial u$, the denominators are never zero for real-valued $q$ and $u$.

Equations~\ref{dlogLdq_nomargin.eqn} are \ref{dlogLdu_nomargin.eqn} are both equal to zero if each of the frequency channels satisfies either $I_{\mathrm{obs},i} + q\,Q_{\mathrm{derot},i} + u\,U_{\mathrm{derot},i} = 0 $ or simultaneously the following two equations
\begin{eqnarray}
\begin{cases}
-q\left(I_{\mathrm{obs},i} + u\,U_{\mathrm{derot},i}\right) + (1+u^2)\,Q_{\mathrm{derot},i} = 0\\
-(1+q^2)\,U_{\mathrm{derot},i} + u\,\left(I_{\mathrm{obs},i} + q\,Q_{\mathrm{derot},i} \right) = 0\, .
\end{cases}
\end{eqnarray}
Since RM synthesis involves sums over frequency channels, while these solutions do not, we do not consider these solutions to be the equivalents of RM synthesis that we are looking for. 
Equation~\ref{dlogLdq_nomargin.eqn} is a polynomial of degree 2 in $q$, and one can easily solve for $q$ as a function of $u$. 
Inserting either one of the solutions for $q(u)$ into the equation for $\partial \log \Lambda /\partial u$ leads to a fraction with in its numerator a polynomial of degree six in $u$ and in its denominator a polynomial of degree eight.
The Abel-Ruffini theorem states that there is no general algebraic solution for polynomials of degree five or higher with arbitrary coefficients \citep{jacobson2009}; indeed, we did not find an obvious solution for $u$ in a reasonable amount of time, not even if we used the computer algebra system $\textsc{Wolfram\,Mathematica}$.
Based on this result we believe that also the more general situation ($\eta\,\ne\,1$ or arbitrary noise variances) cannot be solved algebraically.

\subsection{Marginalize over $I_\mathrm{mod,i}$}\label{marginalization.sec}
Marginalizing over the nuisance parameters $I_\mathrm{mod,i}$ for all frequency channels leads to the new (marginal) likelihood
\begin{eqnarray}
\log{\Lambda'} & = & 
-\frac{1}{2}\sum_{i=1}^{N_\mathrm{ch}} \left[\mathrm{fac}_{1,i} + 2\log\left(\mathrm{fac}_{2,i}\right)\right] \nonumber\\
 & & - \sum_{i=1}^{N_\mathrm{ch}} \left[ \log\left(\sigma_{Q,i}\right) 
 + \log\left(\sigma_{U,i}\right) 
 + \log\left(\sigma_{I,i}\right) \right] \nonumber\\
 & & -N_\mathrm{ch}\left[\log\left(2\pi\right)+2\log\left(\eta\right)\right]\, ,
\label{loglikelihood_prime.eqn}
\end{eqnarray}
where
\begin{eqnarray}
\mathrm{fac}_{1,i} & = & 
\left[\left(Q_{\mathrm{obs},i}-\alpha_i I_{\mathrm{obs},i}\right)^2\sigma_{U,i}^2  \right. \nonumber\\
 & & \left. +\left(U_{\mathrm{obs},i}-\beta_i I_{\mathrm{obs},i}\right)^2\sigma_{Q,i}^2  \right. \nonumber\\
 & & \left. +\left(\beta_i Q_{\mathrm{obs},i} -  \alpha_i U_{\mathrm{obs},i} \right)^2\left(\sigma_{I,i}/\eta\right)^2 \right]/\zeta_i^2\, , \nonumber\\
\begin{pmatrix} \alpha_i \\ \beta_i
\end{pmatrix}
 & = &
\begin{pmatrix} c_i & -s_i \\ s_i & c_i\\
\end{pmatrix}
\begin{pmatrix} q \\ u
\end{pmatrix}\, ,
\nonumber\\
\zeta_i^2 & = & \alpha_i^2\sigma_{I,i}^2\sigma_{U,i}^2 + \beta_i^2\sigma_{I,i}^2\sigma_{Q,i}^2 + \eta^2\sigma_{Q,i}^2\sigma_{U,i}^2\, , \mathrm{and} \nonumber\\
\mathrm{fac}_{2,i} & = & \sqrt{
\left(\frac{1}{\sigma_{I,i}}\right)^2 + 
\left(\frac{\alpha_i}{\eta\,\sigma_{Q,i}}\right)^2 + 
\left(\frac{\beta_i}{\eta\,\sigma_{U,i}}\right)^2\, .  \nonumber
}
\end{eqnarray}
The free parameters $q, u,$ and $\eta$ occur in $\mathrm{fac}_{1,i}$, $\zeta_i$, and $\mathrm{fac}_{2,i}$, therefore, taking the derivative of equation~\ref{loglikelihood_prime.eqn} with respect to these parameters in order to maximize the likelihood leads to non-linear equations in these parameters. 
We tested if analytical solutions can be found that maximize the likelihood if we make the same two simplifying assumptions as in Section~\ref{exact_noise_variances.sec}, i.e., $\eta=1$ and $\sigma_{Q,i}=\sigma_{U,i}=\sigma_{I,i}\equiv\sigma\,\left(\mathrm{constant}\right)$.
Taking the derivative of equation~\ref{loglikelihood_prime.eqn} with respect to $q$ and $u$ leads to two fractions with in their denominators $\sigma^2\left(q^2+u^2+1\right)^2$ and in their numerators
\begin{eqnarray}
\lefteqn{\left(1-q^2+u^2\right)\left(\mathcal{QI}+u\,\mathcal{QU}\right) + qu^2\left(\mathcal{QQ}-\mathcal{UU}-N_\mathrm{ch}\sigma^2\right)} \nonumber\\
 & & - 2qu\,\mathcal{UI} - q\left(\mathcal{II}-\mathcal{QQ}_\mathrm{obs}-\mathcal{UU}_\mathrm{obs}-N_\mathrm{ch}\sigma^2-\mathcal{UU}\right) \nonumber\\
 & & -q^3N_\mathrm{ch}\sigma^2\,  \label{dlogLdq_margin.eqn} \\	
\lefteqn{\left(1+q^2-u^2\right)\left(\mathcal{UI}+q\,\mathcal{QU}\right) + q^2u\left(-\mathcal{QQ}+\mathcal{UU}-N_\mathrm{ch}\sigma^2\right)} \nonumber\\
 & & - 2qu\,\mathcal{QI} - u\left(\mathcal{II}-\mathcal{QQ}_\mathrm{obs}-\mathcal{UU}_\mathrm{obs}+N_\mathrm{ch}\sigma^2+\mathcal{QQ}\right) \nonumber\\
 & & -u^3N_\mathrm{ch}\sigma^2\, ,
\label{dlogLdu_margin.eqn}
\end{eqnarray}
where 
\begin{eqnarray}
\mathcal{QQ}_\mathrm{obs} & = & \sum_{i=1}^{N_\mathrm{ch}} Q_{\mathrm{obs},i}^2  \\
\mathcal{UU}_\mathrm{obs} & = & \sum_{i=1}^{N_\mathrm{ch}} U_{\mathrm{obs},i}^2\, .
\end{eqnarray}
Equations~\ref{dlogLdq_margin.eqn} and \ref{dlogLdu_margin.eqn} are mixed polynomials of degree three in both $q$ and $u$.
If we insert each of the three roots $q(u)$ of the numerator of $\partial \log \Lambda /\partial q$ into the expression for $\partial \log \Lambda /\partial u$ this leads to a fraction with a polynomial of degree seven in $u$ in its numerator and a polynomial of degree eight in its denominator. 
We did not find solutions for $q$ and $u$ in a reasonable amount of time.

\section{Deriving an approximation to the log likelihood if $L \propto I$}\label{Appendix_B}
To derive equation~\ref{loglikelihood_prime_simplified.eqn} from equation~\ref{loglikelihood_prime.eqn} we assume that the noise variances in Stokes $Q$ and $U$ are equal for each frequency channel but are allowed to vary between channels. We introduce $\sigma_{L,i}^2 \equiv\sigma_{Q,i}^2 = \sigma_{U,i}^2$, $p^2 \equiv q^2+u^2$, $L_{\mathrm{mod},i} \equiv \sqrt{Q_{\mathrm{mod},i}^2 + U_{\mathrm{mod},i}^2}$, and $L_{\mathrm{obs},i} \equiv \sqrt{Q_{\mathrm{obs},i}^2 + U_{\mathrm{obs},i}^2}$.
Then
\begin{eqnarray}
\alpha_i^2 + \beta_i^2 & = & p^2 = \left(\frac{L_{\mathrm{mod}}}{I_{\mathrm{mod}}}\right)^2 \\
\sigma_{I,i}\,\mathrm{fac}_{2,i} & = & \sqrt{1 + \left(\frac{\sigma_{I,i}}{\eta\sigma_{L,i}}\right)^2\left(\frac{L_{\mathrm{mod}}}{I_{\mathrm{mod}}}\right)^2}  
\end{eqnarray}
\begin{eqnarray}
\lefteqn{\mathrm{fac}_{1,i} = } \nonumber\\
& & 
  \left(\frac{Q_{\mathrm{obs},i} - \alpha_i\,I_{\mathrm{obs},i}}{\eta\,\sigma_{L,i}}\right)^2 + 
  \left(\frac{U_{\mathrm{obs},i} - \beta_i\,I_{\mathrm{obs},i}}{\eta\,\sigma_{L,i}}\right)^2   \nonumber\\
& & 
  + \left(p^2\,I_{\mathrm{obs},i}^2 - 2\left(Q_{\mathrm{obs},i}\alpha_i + U_{\mathrm{obs},i}\beta_i\right)I_{\mathrm{obs},i} + L_{\mathrm{obs},i}^2\right) \nonumber\\
& & 
  \times \left(\frac{1}{p^2\sigma_{I,i}^2+\eta^2\sigma_{L,i}^2} - \frac{1}{\eta^2\sigma_{L,i}^2}\right) \nonumber\\
& &  
  + \left(\frac{\beta_i Q_{\mathrm{obs},i} -  \alpha_i U_{\mathrm{obs},i}}{\eta\,\sigma_{L,i}}\right)^2\frac{\sigma_{I,i}^2}{p^2\sigma_{I,i}^2 + \eta^2\sigma_{L,i}^2}
\, . 
\end{eqnarray}
If the source is weakly polarized ($L_{\mathrm{mod}} \ll I_{\mathrm{mod}}$) and $\sigma_{I,i}~\approx~\eta\,\sigma_{L,i}$ then $p^2$ is approximately zero, and $\sigma_{I,i}\,\mathrm{fac}_{2,i}~\approx~1$, therefore 
$\log{\left(\sigma_{I,i}\,\mathrm{fac}_{2,i}\right)}\approx0$.
Furthermore, since
\begin{eqnarray}
\mathrm{fac}_{1,i} & \approx &
\left(\frac{Q_{\mathrm{obs},i} - \alpha_i\,I_{\mathrm{obs},i}}{\eta\,\sigma_{L,i}}\right)^2 + 
\left(\frac{U_{\mathrm{obs},i} - \beta_i\,I_{\mathrm{obs},i}}{\eta\,\sigma_{L,i}}\right)^2  \nonumber \\
 & & 
 + \left(\frac{\beta_i Q_{\mathrm{obs},i} -  \alpha_i U_{\mathrm{obs},i}}{\eta\,\sigma_{L,i}}\right)^2\, ,
\end{eqnarray}
equation~\ref{loglikelihood_prime.eqn} simplifies to equation~\ref{loglikelihood_prime_simplified.eqn}.

\end{document}